\documentclass[a4paper,11pt]{article}

\usepackage[margin=1.14in]{geometry} 
\usepackage{tabularx}
\usepackage{multirow}
\usepackage{arydshln}
\usepackage{amsmath}
\usepackage{amsthm}
\usepackage{amssymb}
\usepackage{graphicx}
\usepackage{pdfpages}
\makeatletter
\makeatletter \renewcommand{\@dotsep}{10000} \makeatother
\usepackage{wrapfig}
\usepackage[utf8]{inputenc}
\usepackage{lmodern}
\usepackage{hyperref}
\hypersetup{
	unicode,
	colorlinks,
	breaklinks,
	urlcolor=cyan, 
    linkcolor=black, 
	pdfauthor={Author One, Author Two, Author Three},
	pdfproducer={LaTeX},
	pdfcreator={pdflatex}
}
\usepackage[T1]{fontenc} 
\usepackage{amsmath}

\newcommand{\be}{\begin{eqnarray}}
\newcommand{\ee}{\end{eqnarray}}
\def\be{\begin{equation}}
\def\ee{\end{equation}}
\def\bea{\begin{eqnarray}}
\def\eea{\end{eqnarray}}

\newcommand{\gsim}{\;\raisebox{-0.9ex}{$\textstyle\stackrel{\textstyle >}{\sim}$}\;}
\newcommand{\lsim}{\;\raisebox{-0.9ex}{$\textstyle\stackrel{\textstyle<}{\sim}$}\;}
\def\lsim{\raise0.3ex\hbox{$\;<$\kern-0.75em\raise-1.1ex\hbox{$\sim\;$}}}
\def\gsim{\raise0.3ex\hbox{$\;>$\kern-0.75em\raise-1.1ex\hbox{$\sim\;$}}}

\usepackage{graphics}

\usepackage{epsfig}
\usepackage{slashed}
\usepackage[utf8]{inputenc}
\usepackage{multirow}
\usepackage{pstricks}
\usepackage{dcolumn}

\usepackage[sort&compress,numbers,square]{natbib}

\theoremstyle{plain}

\theoremstyle{definition}

\usepackage{graphicx, color}

\title{Sharpening the $A\to Z^{(*)}h $ Signature of the Type-II 2HDM \\[0.25cm]
at the LHC through Advanced Machine Learning}

\author{\Large{W. Esmail$^a$, A. Hammad$^{b,c}$ and  S. Moretti$^{d,e}$}}

\date{
\small{
$^a$ GSI Helmholtzzentrum f\"{u}r Schwerionenforschung GmbH, 64291 Darmstadt, Germany. \\ 
$^b$ Institute of Convergence Fundamental Studies, Seoul National University of
Science and Technology, 232 Gongneung-ro, Nowon-gu, Seoul, 01811, Korea.\\ 
$^c$ Theory Center, IPNS, KEK, 1-1 Oho, Tsukuba, Ibaraki 305-0801, Japan.\\
$^d$ School of Physics and Astronomy, University of Southampton, Highfield,
Southampton, UK.\\
$^e$ Department of Physics $\&$ Astronomy, Uppsala University, Box 516, SE-751 20 Uppsala, Sweden.}
}
\begin{document}
	\maketitle
	\vspace{4mm}
	\begin{abstract}
 \normalsize{The $A\to Z^{(*)}h$ decay signature has been highlighted as possibly being the first testable probe of the Standard Model (SM) Higgs boson discovered in 2012 ($h$) interacting with Higgs companion states, such as those existing in a 2-Higgs Doublet Model (2HDM), chiefly, a CP-odd one ($A$). The production mechanism of the latter at the Large Hadron Collider (LHC) takes place via $b\bar b$-annihilation and/or $gg$-fusion, depending on the 2HDM parameters, in turn dictated by the Yukawa structure of this Beyond the SM (BSM) scenario. Among the possible incarnations of the 2HDM, we test here the so-called Type-II, for a twofold reason. On the one hand, it intriguingly offers two very distinct parameter regions compliant with the SM-like Higgs measurements, i.e., where the so-called `SM limit' of the 2HDM can be achieved. On the other hand, in both configurations, the $AZh$ coupling is generally small, hence the signal is strongly polluted by backgrounds, so that the exploitation of Machine Learning (ML) techniques becomes extremely useful. In this paper, we show that the application of advanced ML implementations can be decisive in establishing such a signal. 
 This is true for all distinctive kinematical configurations involving the  $A\to Z^{(*)}h$ decay, i.e., below threshold ($m_A<m_Z+m_h$), at its maximum ($m_Z+m_h<m_A<2m_t$) and near the onset of $t\bar t$ pair production ($m_A \approx 2m_t$), for which we propose Benchmark Points (BPs) for future phenomenological analyses. }
\end{abstract}
\newpage
\noindent\rule{\textwidth}{1pt}
\tableofcontents
\noindent\rule{\textwidth}{0.2pt}
\maketitle \flushbottom
\vspace{4mm}
\section{Introduction}
\label{sec:intro}
\subsection{The physics problem}

The Higgs boson discovered at the LHC in 2012 by both the ATLAS and CMS collaborations \cite{Aad:2012tfa,Chatrchyan:2012xdj} is very much consistent with the one embedded in the SM, when it comes to the innumerable measurements performed of its properties (i.e., mass, Yukawa and gauge couplings, spin, CP quantum numbers). Yet, the so-called SM limit, whereby a Higgs boson state of an enlarged Higgs sector can play the role of the SM one, exists in a variety of BSM scenarios. 

Amongst the latter, we concentrate here on the 2HDM \cite{Branco:2011iw}, being the simplest BSM realization of the Higgs mechanism of Electro-Weak Symmetry Breaking (EWSB) employing  the only Higgs field multiplet structure revealed by Nature so far, i.e., the doublet one.  Such a BSM scenario is rather varied in its Higgs sector as, 
after EWSB, it contains five physical Higgs states. These are the $A$  (massive, neutral and CP-odd), the $H^\pm$ (massive, charged and with mixed CP) states alongside two massive neutral CP-even ones, $h$ and $H$ (with, conventionally,  $m_H>m_h$). Either of the latter two can be the aforementioned SM-like Higgs state with a mass of 125 GeV or so, i.e., the $h$ (in which case one speaks of a `normal mass hierarchy' scenario) or the $H$ (in which case one speaks of an `inverted mass  hierarchy' scenario). 
Herein, we assume the first configuration, such that $m_h=125$ GeV (with all $h$ couplings being SM-like). 
A 2HDM is phenomenologically appealing also for another reason, it can   easily comply with the strong limits from EW Precision Observables (EWPOs), as it suffices to set the $H^\pm$ mass  somewhat degenerate with those of the $A$ and/or $h/H$ states. 
Finally, the 2HDM can dispense of large Flavour Changing Neutral Currents (FCNCs) by simply invoking a $\mathbb{Z}_2$ symmetry between the two Higgs doublet fields, which can prevent these from occurring at tree-level. In turn, this implies well-defined Yukawa structures (which we will describe in detail below), depending on how the two Higgs doublets couple to  fermions, which go under the name of Type-I, -II, lepton-specific and flipped \cite{Branco:2011iw}. 

Amongst all these, we concentrate here on the Type-II case. We do so as this realization of the 2HDM is the most challenging one phenomenologically. In fact, it  implies a lower bound on the charged Higgs mass around 600 GeV, as per constraints coming from $b\to s\gamma$ transitions \cite{Misiak:2017bgg}. As mentioned, the EWPOs then require also the $A$ and/or $H$ states to be rather heavy. In fact, in the 2HDM Type-II, two different regions over the $(\cos (\beta - \alpha), \tan \beta)$ plane\footnote{Here, $\alpha$ is the mixing angle between the $h$ and $H$ states and $\tan\beta$ is the ratio of the Vacuum Expectation Values (VEVs) of the two doublets.} can realize the aforementioned SM-like configuration (see, e.g., Refs.~\cite{Ferreira:2014naa,Bernon:2015wef,Basler:2017nzu,Ferreira:2017bnx}). According to the analysis of Refs.~\cite{Accomando:2019jrb,Accomando:2022nfc}
(see also Ref.~\cite{Bernon:2015wef}), in the first one, the so-called `alignment limit', whereby $\cos(\beta-\alpha)\to 0$ (and the couplings of the $h$ state to $u$- and $d$-type quarks have the same sign as those in the SM) the CP-odd Higgs state is required to be rather heavy ($m_A \ge 350$ GeV) while, in the second one, the so-called `wrong-sign scenario', whereby $\cos(\beta-\alpha)$ can reach 0.4 or so (and the couplings of the $h$ state to $u$($d$)-type quarks have the same(opposite) sign as(to) those in the SM)   the $A$ mass can be as light as 200 GeV or so (a mass region recently explored 
in Refs.  \cite{Accomando:2020vbo,Akeroyd:2023kek}).

Thus, the 2HDM Type-II offers the possibility to LHC searches of establishing sensitivity to the presence of the $A$ state over a wide mass range. The latter is  most copiously produced via $b\bar b,gg\to A$ and can, in particular, decay via $A\to Z^{(*)}h$, which can then altogether be elevated to a new $A$ search channel, alongside the traditional ones in $\tau^+\tau^-$ (for $m_A<2m_t$) and $t\bar t$ (for $m_A>2m_t$) final states, since the $h$ mass is now known rather precisely. The importance of the $A\to Z^{(*)}h$ decay channel has been  repeatedly emphasised in literature, as it would simultaneously allow one to establish the presence of an extended Higgs sector as well as the gauge structure of the theory embedding it.  

It is the purpose of our paper the one of proposing new searches for the $b\bar b,gg\to A\to Z^{(*)}h\to l^+l^-b\bar b$ channel at the LHC over an extended $m_A$ range, from values both below $m_A+m_Z$ (where the neutral weak gauge boson is off-shell, $Z^*$) and (far) above it (where the neutral weak gauge boson is on-shell, $Z$). Furthermore, in the light of the fact that the $AZh$ vertex is suppressed in the 2HDM Type-II so that SM backgrounds to the aforementioned $l^+l^-b\bar b$ signature are significant, in order to establish sensitivity to this BSM scenario too, we deploy here advanced Machine Learning (ML) methods that could well be adopted by ATLAS and CMS in searching for this 2HDM signal, as they surpass the state-of-the-art used so far in the corresponding analyses. 

\subsection{The ML algorithms}

{Specifically, we carry out our search by utilizing a set of Deep Neural Networks (DNNs) that span all data types at the LHC, e.g., kinematical  distributions, energy deposit of charged hadrons and (reconstructed) four-momenta of final state particles. A Multi-Layer Perceptron (MLP) network that analyzes the constructed kinematical distributions of the final state particles is also used. Then, a Convolution Neural Network (CNN), which analyzes jet images that can be constructed by visualising the $pT$ (transverse momentum) distributions of the final state jets is exploited. Furthermore, we adopt a new method for a Siamese Neural Network (SNN) which is a twin encoder model with two training stages. The model maps the high dimensional input feature space to lower dimensional space (latent space) such that the Euclidean distance between images from different classes is maximal. For this purpose the SNN minimizes a modified contrastive loss function in the first training stage, while in the second training stage it minimizes an entropy loss function. Also, we adjust a Hybrid Deep Neural Network (HDNN), which is a two streams input network that can analyze the kinematical distributions and constructed jet images at the same time. Finally, to tackle the issue of sparse pixels in jet images, we utilize a suite of Graph Neural Networks (GNNs) to examine the graphs developed from the four-momenta of the final state particles. In this scenario, we employ four distinct GNNs: a Dynamic Graph Convolution Neural Network (DGCNN), a Graph Convolution Network (GCN), a Graph Attention (GAT) network, and a Graph Sample and AggreGate (GraphSAGE) network. \par

In order to study the influence of each network individually, we utilize a linear CKA to evaluate the similarity among hidden layer representations. This approach is necessary as Deep Learning (DL) models are typically considered as computational tools lacking an interpretative element, i.e.,  without innate explanations for their results. Instead, we leverage CKA to analyze the information learned by the hidden layers of each model, providing a robust explanation for each model's individual classification accuracy. Despite the linear CKA's innate ability to explain the reported model accuracy by scrutinizing the representation pattern within a model's hidden layers, we also use the CKA to compare the classification accuracy among different used models.}

The motivation for our comparative analysis of such ML tools is the one of establishing the best one to use in order to extract the aforementioned 2HDM signal, also in the light of the fact that ATLAS and CMS have recently started exploiting advanced ML algorithms, generally based on  graph and transformer networks, so that our phenomenological study can serve the purpose of validating such approaches (mainly used for jet analyses) also in the specific context of 2HDM physics. 

\subsection{Layout}

Our paper is organised as follows. In the next section, we describe the 2HDM, in particular, its Type-II realization. We then illustrate our overall analysis strategy, followed by a detailed description of the various ML methods that we advocate. {Then we provide an analysis of the learned representations of the hidden layers of each model thereby offering a robust explanation of the reported accuracy of our results.} After which, we present the latter and finally conclude.

\section{The 2HDM}
\label{sec:The model}
In this section we first give a brief review of the 2HDM with type-II Yukawa couplings focusing  on the aspects of it which are relevant to our analysis. We then describe theoretical and experimental constraints applicable to it. We finally scan over its parameter space to extract interesting BPs to be used in our numerical analysis. 
\subsection{The Higgs potential}
\label{sec:themodel}
The  2HDM is an extension of the SM  through a second $SU(2)_L$ Higgs doublet with the same quantum numbers under the SM symmetry gauge group \cite{Branco:2011iw,Lee:1973iz}. The two $(SU)_L$ doublet fields, $\phi_1$ and $\phi_2$, are defined as
\begin{equation}
	\phi_1 = \begin{pmatrix}
	\eta_1^+ \\
	(v_1 + h_1 + i h_3)/\sqrt{2} \\
	\end{pmatrix},\qquad
	\phi_2 = \begin{pmatrix}
	\eta_2^+ \\
	(v_2 + h_2 + i h_4)/\sqrt{2} \\
	\end{pmatrix}\,,
\end{equation}%
in terms of four (pseudo)real scalar fields $h_i$, with $i =1,\ldots, 4$, two complex charged fields $\eta_i^+$, with $i=1,2$, and two Vacuum Expectation Values (VEVs) $v_i$, with $i=1,2$.
The Lagrangian density of the model can be decomposed as 
\begin{equation}
    \mathcal{L}_{\rm 2HDM} = \mathcal{L}_{\rm SM} + \mathcal{L}_{\phi}+V_\phi+Y_\phi\,,
\end{equation}
where $\mathcal{L}_{\rm SM}$ contains the kinetic terms for the SM gauge fields and fermions, $\mathcal{L}_{\phi}$ contains those of the two Higgs doublet fields, $V_\phi$ denotes the Higgs potential of the two doublet fields and $Y_\phi$ is the Yukawa part which gives rise to the couplings between the Higgs fields and SM fermions. 
The most general 2HDM Higgs potential is given by
\begin{equation}
\begin{split}
    V_\phi &= m^2_{11} (\phi^\dagger_1 \phi_1) + m^2_{22} (\phi^\dagger_2 \phi_2) - \left[m^2_{12}(\phi^\dagger_1\phi_2)+\text{h.c.}\right]\\
     &+\lambda_1 (\phi^\dagger_1 \phi_1)^2+\lambda_2 (\phi^\dagger_2 \phi_2)^2  +\lambda_3 (\phi^\dagger_1 \phi_1) (\phi^\dagger_2 \phi_2) +\lambda_4 (\phi^\dagger_1 \phi_2) (\phi^\dagger_2 \phi_1)  \\
      &+\frac{1}{2}\left[ \lambda_5 (\phi^\dagger_1\phi_2)^2+ \left[\lambda_6(\phi^\dagger_1\phi_1)+ \lambda_7(\phi^\dagger_2\phi_2)\right](\phi^\dagger_1\phi_2)+\text{H.c.}\right]  \,.
\end{split}
\label{eq:Vphi}
\end{equation}

Such a   potential  allows for Flavor Changing Neutral Currents (FCNCs) at tree level, though,  which are strongly constrained by experimental measurements. Adding a global $Z_2$ symmetry to the potential, with $(\phi_1,\phi_2)\to(\phi_1,-\phi_2)$ transformations,  prevents the existence of FCNC sources in it \cite{Glashow:1976nt}. However, the most general Yukawa interaction violates such a $Z_2$ symmetry, thus leading again to potentially FCNCs at tree level \cite{Ginzburg:2004vp}. 
Thus, to tame the latter, only specific Yukawa structures, known as the aforementioned Types \cite{Branco:2011iw}, are allowed. However, to enable EWSB compliant with the measured particle spectrum of the SM, a softly broken $Z_2$ symmetry should be enabled, by requiring a small but non-vanishing mass $m^2_{12}(\phi^\dagger_1\phi_2)$ and setting $\lambda_6=\lambda_7=0$. (Herein, softly means that the model still respects the $Z_2$ symmetry at small distances in all order of perturbation theory.)  The `soft' mass $m^2_{12}$ and $\lambda_5$ are in general complex, though,  with two phases $m^2_{12} = |m^2_{12}|e^{i\eta(m^2_{12})}$ and $\lambda_5 = |\lambda_5|e^{i\eta(\lambda_5)}$ \cite{Antusch:2020ngh,Antusch:2021oit}. In the following, we will consider a real potential that preserves the CP symmetry, thus  with vanishing complex phases, $\eta(m^2_{12})=\eta(\lambda_5)=0$. In such a configuration of the 2HDM, then 7 independent parameters remain, which are $\lambda_i$, with $i=1,\dots 5$, $\tan\beta=v_2/v_1$ and $m_{12}^2$, from which the physical parameters, i.e., Higgs boson masses and couplings, are obtained, with the constraint that one of the former must be set to 125 GeV or so (which in our case is the one of the $h$ field). Finally, as mentioned already, amongst the possible Yukawa structures, we restrict our study to the Type-II only.

The tree level mass matrix squared for the Higgs fields can be obtained as
\begin{equation}
    \left(\mathcal{M}^2 \right)_{ij}= \frac{\partial V_\phi}{\partial h_i\partial h_j} \Bigg|_{h_{i,j}=0}\,,
\end{equation}
where the $h_i$'s ($i=1,\ldots ,4$) are the four components of the complex doublet fields. Upon EWSB, three physical neutral scalars are obtained after diagonalizing the corresponding mass matrices, two CP-even (scalar) ones ($h, H$) and a CP-odd (pseudoscalar) one ($A$),  with masses given by
\begin{align}
    \label{eq:massh1h2}
    m^2_{h,H} & = \frac{1}{2}\left[\chi^2_{11}+\chi^2_{22}\mp \sqrt{(\chi^2_{11}-\chi^2_{22})^2+4(\chi^2_{12})^2}   \right]\,, \\
    \label{eq:masshA}
    m^2_A  &= \frac{2m^2_{12}}{\sin2\beta}-\lambda_5 v^2\,,
\end{align}
with
\begin{align}
    \label{eq:mass11}
    \chi^2_{11} & =m^2_{12}\tan\beta+2\lambda_1 v^2\cos^2\beta\,,\\
    \label{eq:mass22}
    \chi^2_{22} & =m^2_{12}\cot\beta+2\lambda_2 v^2\sin^2\beta\,,\\
    \label{eq:mass12}
    \chi^2_{12} & =-m^2_{12}+\frac{1}{2}(\lambda_3+\lambda_4+\lambda_5) v^2\sin2\beta\,,
\end{align}
where the VEVs satisfy the relation $v=\sqrt{v_1+v_2}$ (with $v$ being the SM one)\footnote{The other two Higgs states emerging from the 2HDM after EWSB are charged and are denoted by $H^\pm$.}. As intimated, in the following, we will consider $h$ as the SM-like Higgs boson discovered at the LHC in 2012, with . 

To stay with the neutral Higgs sector, the imposed CP conservation only allows for tree level couplings between two massive gauge bosons and the CP-even Higgs states while the CP-odd Higgs state can only couple to a gauge boson and a CP-even Higgs one. Furthermore, all neutral Higgs states can couple to fermions. The couplings strength of the neutral Higgs bosons to both matter and forces are parameterized in terms of $\tan\beta$ and another parameter, $\alpha$, which is the mixing angle between the CP-even Higgs states \cite{Branco:2011iw}. Specifically, the coupling strength  of the $AZh$ vertex is proportional to $\cos(\beta-\alpha)$.  
\subsection{Constraining the 2HDM free parameters}
\label{sec:addconst}
The 2HDM free parameters are constrained from various theoretical considerations and experimental observations. In order to account for the  perturbativity of the Higgs  potential, the magnitude of the couplings in the Higgs potential is constrained to $|\lambda_i|\le 4\pi$ ($i=1, \dots 5$). The stability of the model vacuum constrains a combination of these couplings, as follows \cite{Ivanov:2015nea}
\begin{equation}
\lambda_1,\lambda_2 > 0,\hspace{4mm} \lambda_3+\sqrt{\lambda_1\lambda_2}> 0, \hspace{4mm} \lambda_3+\lambda_4-\lambda_5+\sqrt{\lambda_1\lambda_2}> 0\,. 
\end{equation}
The contribution of the 2HDM particles to EWPOs at the loop level affects the measured oblique parameters, which are constrained from  global fits to be \cite{Baak:2012kk}
\begin{equation}
S = 0.03 \pm 0.10 ,\hspace{4mm} T= 0.05 \pm 0.12 , \hspace{4mm} U = 0.03 \pm 0.10 \,,
\end{equation}
so that we account for these limits too. The precise measurements of the SM Higgs mass and coupling strengths by the ATLAS and CMS experiments add extra bounds on the properties of the SM-like Higgs, $h$ \cite{CMS:2012qbp,ATLAS:2016neq,
ATLAS:2015yey,ATLAS:2012yve}. Furthermore, the other neutral and charged Higgs states undergo  constraints from null resonance searches at various colliders, see, e.g.,  \cite{LEPHiggsWorkingGroupforHiggsbosonsearches:2001rfu,CDF:2008ghn,ATLAS:2019old,CMS:2019bnu}.
The contribution of the charged Higgs boson to $B$ meson decays sets severe bounds on the $(m_{H^\pm},\tan\beta)$ plane, as mentioned. The dominant bounds come from the following Branching ratios (Br) measurements:  
Br$(B^{+}\to \tau^{+}\nu) = (1.06\pm0.19) \times 10^{-4}$ and $Br(B\to S\gamma)_{E_\gamma \ge 1.6\text{GeV}}) = (3.32\pm0.15) \times 10^{-4}$ \cite{HFLAV:2019otj}. Specifically, for large $\tan\beta$ the charged Higgs boson mass is constrained to be $m_{H^\pm}\ge 600$ GeV or so while for $\tan\beta \le 10$ such a mass bound is significantly relaxed \cite{Hammad:2022wpq,Enomoto:2015wbn}.
\subsection{Assessment of the parameter space}
\label{sec:Assessment}
In order to find  viable parameter space points that satisfy all the mentioned constraints we scan over the aforementioned Higgs potential free parameters. For fast convergence we use the ML assisted scanner package of Ref.~\cite{Hammad:2022wpq} to scan over the following ranges:
\begin{equation}
\begin{gathered}
    0\le\lambda_1\le 10,\qquad
  0\le\lambda_2\le 0.2,\qquad
  -10\le\lambda_3\le 10,\qquad
  -10\le\lambda_4\le 10,\\
    -10\le\lambda_5\le 10,\qquad
  1\le\tan\beta\le 45,\qquad
    -6000 \text{ GeV}^2\le m^2_{12}\le 0 \text{ GeV}^2\,.
\end{gathered}
\end{equation}
(The narrow range of $\lambda_2$ is to keep the SM-like Higgs $h$ as the lightest neutral Higgs state.) 
\begin{figure}[tbh!]
    \includegraphics[scale=0.28]{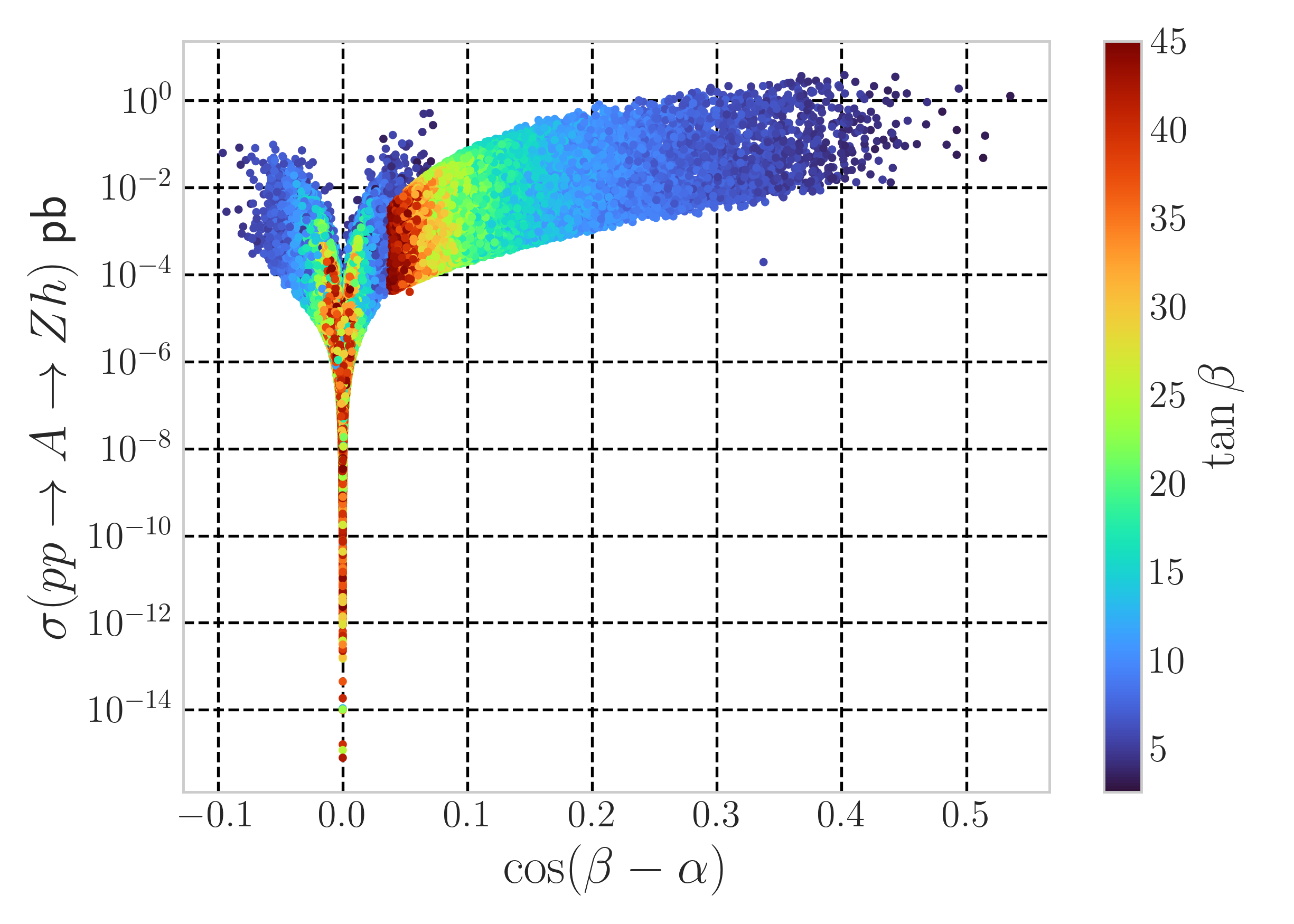}~~
    \includegraphics[scale=0.28]{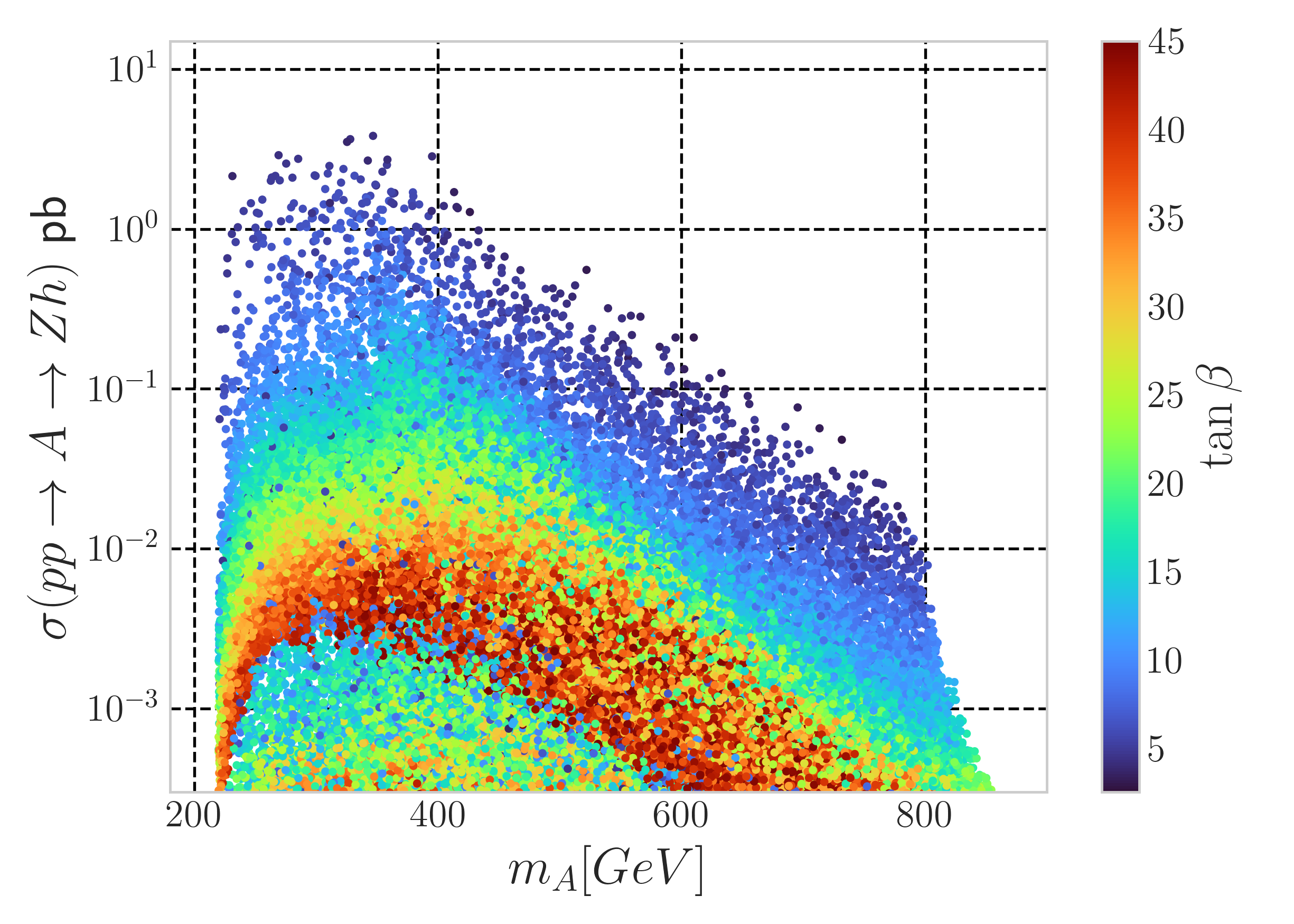}~~\includegraphics[scale=0.28]{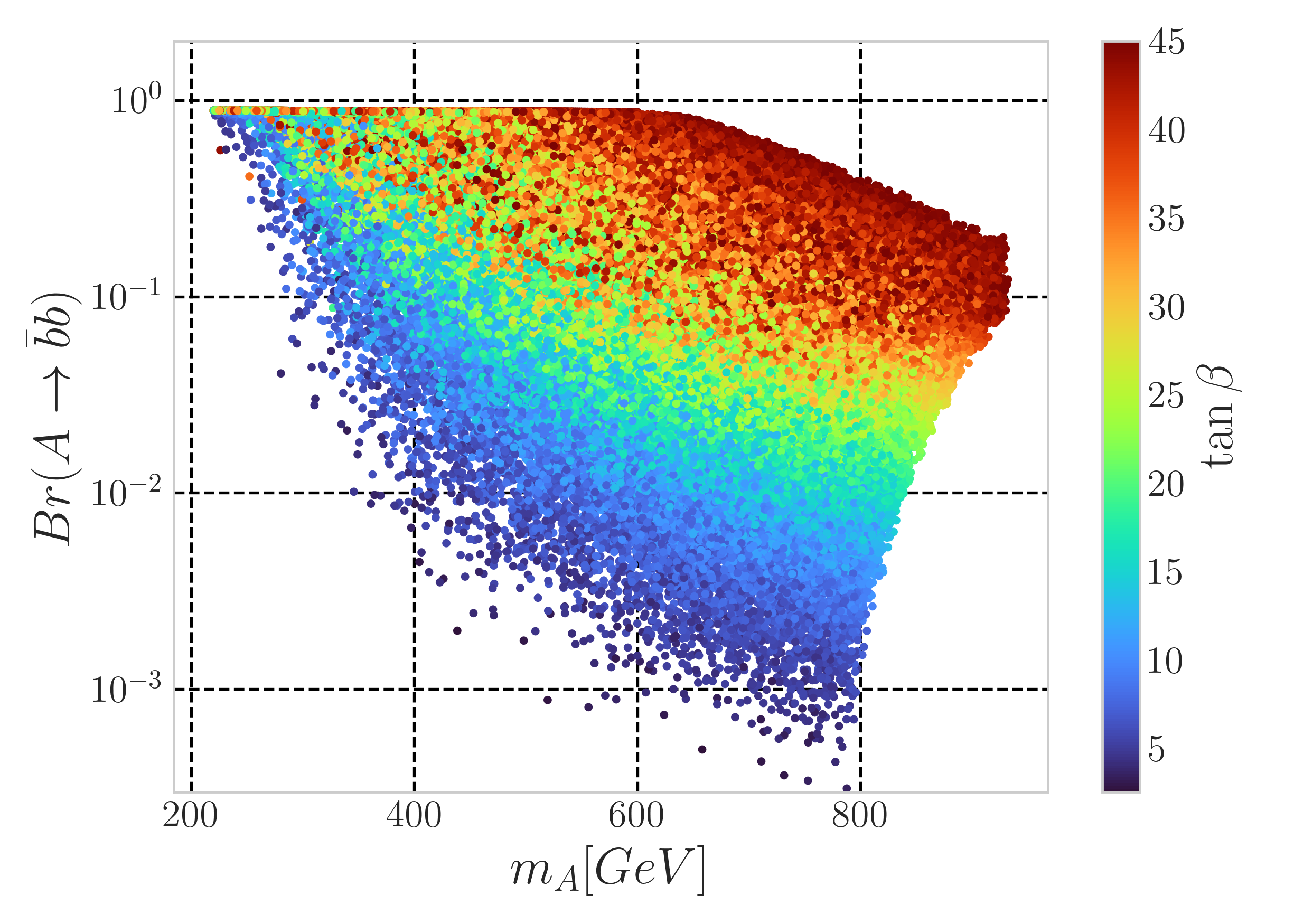}
    \caption{The scan output points that satisfy all constrains. (Left) Total cross section of the process $pp\to A\to Zh$ at $\sqrt{s}= 14$ GeV versus $\cos(\beta-\alpha)$. (Middle) The same versus $m_A$. (Right) The decay rate $Br(A\to b\bar b)$ versus $m_A$. The color bar represents the corresponding $\tan\beta$ value in all plots.}
\label{fig:1}
\end{figure}
As a result, we obtain $300\, 000$ points that satisfy all constraints, which are shown in Fig.~\ref{fig:1}. The left plot shows the total cross section of the process $pp\to A\to Zh$ at $\sqrt{s}= 14$ GeV versus $\cos(\beta-\alpha)$, with the color bar representing the corresponding $\tan\beta$ value. The overall coupling strength is proportional to some function of $\tan\beta$, depending upon the relative size of the $ht\bar t$ and $hb\bar b$ couplings at production level times 
$\cos(\beta-\alpha)$ at decay level, the latter modulated by the functional form of the total width in terms of $\alpha$ and $\beta$. The middle plot shows the same data points
mapped against $m_A$. By combining these two plots, it is clear that the production times decay cross section can be up to $\sim \mathcal{O}(1$ pb) for $m_A\le 400$ GeV and/or $\tan\beta < 10$. The right plot shows the $Br(A\to b\bar b)$, as this is the dominant one over the mass range of interest here, i.e., $m_A < 600$ GeV or so, while for larger $m_A$ the dominant decay modes are $A\to Z^{(*)}H$ and, mostly, $A\to t\bar t$. Indeed, the $A\to Z^{(*)}h$ mode pursued here is never dominant, although it is maximised in the region between $m_Z+m_h$ and $2m_t$. 

As for the separate dynamics of production and decay, it is worth emphasizing the following. On the one hand, for smaller $\tan\beta$, the main contribution to the production cross section comes from $gg\to A$ (i.e., $gg$-fusion) while, for larger $\tan\beta$,  the dominant one is $b\bar b\to A$ (i.e., $b\bar b$-annihilation). On the other hand, for the $A\to Z^{(*)}$ decay rate, the dependence on $\tan\beta$ is less straightforward. As for the further two transitions, $Z^{(*)}\to l^+l^-$ ($l=e,\mu$) and $h\to b\bar b$, these (essentially) are SM processes. Finally, it is worth mentioning that, when the 
top quark loop (entering $gg$-fusion) exhibits an imaginary part for $m_A > 2 m_t$, there occurs a destructive interference of our signal with the $pp\to Z^{(*)}b\bar{b}$ process, which yields a small reduction of the total cross section \cite{Djouadi:2019cbm,Jung:2015etr}, which we neglect here.

\section{Analysis strategy}
\label{sec:intro}
\begin{figure}[h!]
\centering
\includegraphics[scale=0.8]{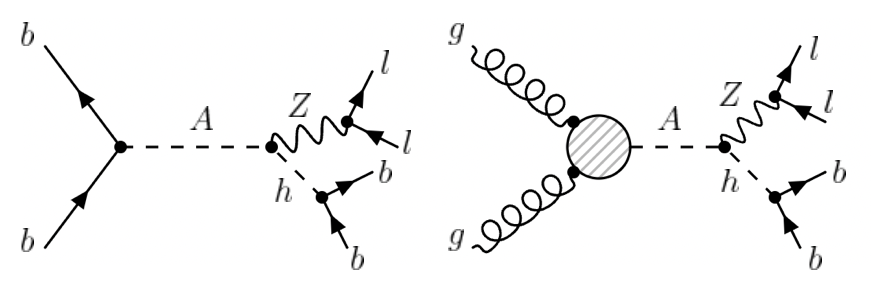}
\caption{Feynman diagrams for the considered signal subprocesses.}
\label{fig:2}
\end{figure}

In this section, we numerically investigate our chosen signature of the CP-odd Higgs boson of the 2HDM Type-II at Run 3 of the LHC and HL-LHC using different recent ML models.  Thus, we concentrate on the  process $pp\to A\to Zh$ with $\sqrt{s}=14$ TeV and an integrated luminosity $L_{\rm int}$ of 300 and $3000$ fb$^{-1}$.
The subprocesses of interest  are initiated by $b\bar b$-annihilation and 
$gg$-fusion, eventually yielding a lepton $l^+l^-$ ($l=e,\mu$) and  $b$-jet pair, as shown in Fig. \ref{fig:2}. 
We first review experimental results on this process obtained by ATLAS and CMS, we then borrow the most essential elements of one of their (kinematical) analyses to introduce our own approach, before moving on to describe the ML part of it.

\subsection{Current ATLAS and CMS results}

There are several searches that have been carried out at the LHC looking for the $A$ state via the $A\to Zh$ decay, i.e., with the $Z$ on-shell (which we will describe in a forthcoming section), typically done  by using   $Z\to l^+l^-$ ($l=e,\mu$) and $h\to b\bar b$ decays. However, these all concentrated  on an $A$ mass range starting from $m_Z+m_h \approx$ 215 GeV, i.e., assuming decays of the CP-odd Higgs boson into $Z$ and $h$ particles both being on-shell. While this assumption is fully justified in the case of the Higgs boson, which has a typical width of order 10 MeV at most (according to latest $h$ measurements at the LHC), it is less so for the gauge boson, for which the width-to-mass ratio is of order $3\%$. {Off-shell effects involving the $Z$ boson are therefore not negligible, hence searching for the CP-odd Higgs boson decaying into $Z^*h$ is of phenomenological interest, as recently advocated in, e.g., \cite{Accomando:2020vbo,Akeroyd:2023kek}.}
Up-to-date searches for $pp\to A\to Zh$ signals at the LHC, in a variety of final states, were recently reviewed in great detail in another paper of one of us \cite{Akeroyd:2023kek}, to which we refer the reader, including their interpretation in two different 2HDM Types. Here, we limit ourselves to the case of the Type-II considered (in normal mass hierarchy) and refers specifically to the $Zh\to l^+l^-b\bar b$ ($l=e,\mu$) signature.

The decay $A\to Z h$ has been searched for at the LHC by both the ATLAS and CMS  collaborations assuming the case of normal mass hierarchy  (i.e., $m_{h}=125$ GeV) and  an on-shell $Z$ boson, i.e., specifically assuming $m_{A} \ge m_Z+m_h$. Just like here, in such searches, the $A$ state is assumed to be produced via $b\bar b, gg\to A$\footnote{Although the emulation of the first subprocess via $gg\to A b\bar b$ is sometimes used \cite{ggvsbb}.} and the decay rates of the $h$ state to fermions are given by the measurements of the $Br$'s of the 125 GeV boson, thus $Br(h\to b\overline b)\approx 57\%$. 

In the CMS analysis of Ref. \cite{CMS:2019qcx} (see also Ref.~\cite{CMS:2019kca}), with $\sqrt s=13$ TeV and 35.9 fb$^{-1}$, targeting the $m_A>225$ GeV range, 
separate searches are carried out for the decays $Z\to e^+e^-$ and $Z\to \mu^+\mu^-$. In each case, the signal is separated into categories with 1, 2 and 3 $b$-jets. In general, the selection efficiencies are similar for the two  production mechanisms in the 1 and 2 $b$-jet categories and they increase slightly with increasing $m_{A}$ while in the 3 $b$-jet category the selection efficiency for $gg\to A b\overline b$ is considerably larger (due to the presence of more $b$-jets in the signal)  than that for $gg\to A$, being almost an order of magnitude greater for $m_{A}<300$ GeV. Furthermore, the SM backgrounds to the $l^+ l^- b\bar b$ signature are largest for the 1 $b$-jet category and smallest for the 3 $b$-jet one. In the 2 $b$-jet category (that we are assume here), the dominant backgrounds are found to be $pp\to Z b\bar b$ and $pp\to t\bar t$, which we shall adopt here too. Searches for the above  signature by the ATLAS collaboration in Refs. \cite{ATLAS:2017xel} (with 36.1 fb${^{-1}}$ of luminosity) and \cite{ATLAS:2020pgp} (with 139 fb${^{-1}}$ of luminosity) have similar strategies and derive comparable limits on the total cross section, over the ranges  $m_{A}>220$ GeV and $m_{A}>280$ GeV, respectively.

{In carrying out our ML driven analysis, we will  compare our results with some of those from the aforementioned ATLAS and CMS analyses, as well as an earlier ATLAS study, the one of Ref.~\cite{TheATLAScollaboration:2016loc}, based on 3.2 fb$^{-1}$ of data and starting from $m_A=220$ GeV, which is, in fact, the one offering the simplest kinematical analysis, upon which we will model ours.}

\subsection{Kinematical analysis}
Following the analysis in \cite{TheATLAScollaboration:2016loc}, we require events to have  two isolated leptons, these being either electrons or muons, and two isolated $b$-tagged jets. The reconstructed events satisfy the following requirements in transverse momentum, pseudorapidity and separation: $pT(l)> 20$ GeV, $pT(j)> 25$ GeV and $\left|\eta(l,\text{jet},b)\right|\le 2.5$. Furthermore, our selection cuts on the invariant masses of the hadronic and leptonic systems are as follows: 75 GeV $< m_{bb}< 145$ GeV and 70 GeV $ < m_{ll}< 110$ GeV.
Jets are reconstructed with the anti-$k_T$ algorithm \cite{Cacciari:2008gp} but our results proved stable against a change of clustering algorithm to the Cambridge-Aachen one \cite{Dokshitzer:1997in,Wobisch:1998wt}.   

Reconstructed events are required to have at least two isolated $b$-jets with cone radius $R=0.4$ using a flat $b$-tagging efficiency of $70\%$. For the mistagging rate of gluon and light quark jets as $b$ ones, we adopt a flat rate of $10^{-3}$  while, for $c$-jets, we use $10^{-2}$.  

The dominant background contributions come from the $W^+W^-$ leptonic decays in the $pp\to  t\bar{t}$ process and from $pp\to Z b\bar{b}$, where the $Z$ boson decays leptonically. Other background processes like single-top production, di-gauge boson production,  associated production of a gauge boson with the SM Higgs and $pp\to W^\pm \bar{b}b$ are not considered as they can be removed by the basic cuts applied here \cite{TheATLAScollaboration:2016loc}.  

\begin{table}[h!]
\caption{Input parameters for our four BPs. The last column shows the total cross section for the process depicted in Fig. \ref{fig:2}.}\vspace*{0.25cm}
\begin{tabular}{|p{0.085\textwidth}|p{0.04\textwidth}|p{0.04\textwidth}|p{0.04\textwidth}|p{0.06\textwidth}|p{0.06\textwidth}|p{0.06\textwidth}|p{0.11\textwidth}|p{0.12\textwidth}|p{0.1\textwidth}|}
\hline
$m_A$[GeV] & $\lambda_1$ & $\lambda_2$ & $\lambda_3$ & $\lambda_4$ & $\lambda_5$ & $\tan\beta$& $m^2_{12}$ GeV$^2$ &$\cos(\beta-\alpha)$  & $\sigma_{\text{tot}}$ [fb] \\ \hline
$200$ & $6.81$&$0.14$&$1.86$&$-0.12$&$-0.31$&$5.02$&$-4260$&$0.37$& $65.8$ \\  \hline
$250$ & $6.12$ & $0.14$ & $1.86$ & $-0.11$ &$-0.81$& $5.01$ &$-4270 $ &$0.38$ &$86.49$  \\  \hline
$300$ & $5.22$ & $0.13$ & $4.00$ & $-1.82$ &$-1.44$& $4.68$ &$-4530$ &$0.35$ &$109.42$  \\  \hline
$350$ & $4.69$ & $0.14$ & $3.85$ & $-1.35$ &$-1.57$& $4.38$ &$-5440$ &$0.34$ &$95.68$  \\  \hline
\end{tabular}
\label{tab:tab_1}
\end{table}

We carry out the analysis for four BPs,  with $m_A = 200,250,~300$ and $350$ GeV. For the first BP, the $Z$ boson is produced off-shell while in other three cases is on-shell. The four BPs are chosen from the output of the scanned points mentioned in section \ref{sec:Assessment}, all of which satisfy all relevant theoretical and experimental bounds. Tab. \ref{tab:tab_1} shows the input parameters for the four BPs, alongside their production times decay cross sections, down to the final state $l^+l^-b\bar b$.
We notice that all our BPs belong to the `wrong-sign scenario', i.e., the right-arm region of Refs.~\cite{Accomando:2019jrb,Accomando:2022nfc}, typically offering larger total cross sections than in the `alignment limit', thereby making these particularly amenable to experimental analyses.

Simulation of the signal and background events proceeds through a chain of sequentially automated steps. For events generation and cross section calculation we use MadGraph \cite{Alwall:2014hca} with its standard generation level cuts  (which do not bias our detector level results). For $gg$-fusion, the  loop  implemented in MadGraph is an effective vertex as described in \cite{Belyaev:2012qa}. SPheno \cite{Porod:2011nf,Porod:2003um} is used to compute the numerical value of such an effective vertex at the Leading Order (LO) in perturbation theory.  PYTHIA \cite{Sjostrand:2006za} is exploited for parton showering, hadronization, heavy flavor decays and for adding the soft underlying event, multi-particle scatterings, etc. FastJet \cite{Cacciari:2011ma} is used for jet clustering. The fast simulation of the ATLAS detector was done with the DELPHES package \cite{deFavereau:2013fsa}. Finally, the  standard ATLAS card is modified to be able to simulate the tracks and energy deposit from the charged  hadrons. 

\section{DNN}
After event simulation, we adopt different types of ML models to analyze different categories of events, kinematical distributions, energy deposits of charged hadrons and (reconstructed) four-momenta of the final state particles.

Starting with high-level kinematical distributions, we adopt a MLP model to optimize the separation  between the signal and background distributions \cite{Guest:2016iqz}. The constructed distributions have unique information about the global structure of the signal and background events, thus the structure of the MLP network, with fully-connected layers, is able to analyze the global features ending up with large classification power between the signal and background events. Although the MLP can achieve high classification performance, the fact that some background distributions have similar kinematical structure to signal ones hinders the overall classification power. However, one can improve the classification performance by applying initial cuts that maximize the signal-to-background yield before feeding the distributions to the MLP. Furthermore, the constructed kinematical spectra exhibit a large correlation among each other and applying a cut on any distribution will, in some cases, affect the structure of all others, aspect which then continues to hinder the classification performance of the MLP. In order to control the global impact of the initial cuts, one has then to decorrelate such a dependence across the kinematical variables via the square-root of the covariance matrix or Gaussian transformation of variables as described  in \cite{Hocker:2007ht}. In the end, although the initial cuts may increase the classification performance, we opted not to apply any thus allowing full freedom to the MLP in finding the optimal classification boundaries.

A second approach is to analyze the charged hadrons by exploiting the fact that, in an unbroken $SU(3)_C$, color is conserved in the QCD interaction and provides different color flow structures for different processes. The structure of the color flow depends on the color nature of mediating particles, e.g., the
radiation pattern within and around $b$-(anti)quark pairs from Higgs boson decays is expected  to be different from the radiation pattern of the same  from $tt$ production or $Zbb$ processes. In order to exploit  the color flow properties to classify signal and background events, one can think of the LHC detector as a giant camera and the streams of hadrons as images. The constructed images are two dimensional arrays in the $(\eta-\phi)$ plane while the pixels size is adjusted to be within the detector response and the pixels are weighted by the sum of the total transverse momentum deposited in the corresponding part in the detector \cite{Komiske:2016rsd,Fraser:2018ieu,Cogan:2014oua,deOliveira:2015xxd}.
We adopt a CNN model to analyze the constructed jet images and output the classification probability for signal and background events. The CNN is constructed by combining two different sets of hidden layers, convolution ones and fully-connected ones. Convolution layers are constructed from filters (kernels) that share their weights locally and hence they are able to capture local information stored into the images while the fully-connected layers are handcrafted to analyze the captured local information ending up with global information about the image by adjusting different  structures of the neurons for signal and background images\cite{Kagan:2020yrm,Chung:2020ysf,Pol:2021iqw,Andrews:2018nwy,DiBello:2020bas,Kim:2022miv}. 

Although the CNN is designed to capture local  information of the jet images, there is no guarantee it can capture some hidden information, e.g., similar or dissimilar local information for images from the same or different classes,  respectively. For this purpose, we introduce a SNN \cite{siamese,Blokland:2021onk}. This  is a two-step training network with twin encoders that share their weights. As a twin convolution encoder model, it processes the images in pairs from the same or different classes. In the first training stage the model learns similar features shared amongst images from the same class,  e.g., pairs of signal or background images, by minimizing the latent space Euclidean distance between them  and maximizing the distance between images from different classes, i.e., pairs from signal and background images. Once the latent space is shaped by separating the Euclidean distance between signal and background images, the second training stage starts by freezing the optimized weights for one encoder and adding a fully-connected layer and one output layer with two output neurons to identify the signal and background events. The construction of the SNN enables it to learn hidden features that are shared among each class. We stress here that contrastive learning incorporate a wide range of supervised and unsupervised DL models such as SimCLR \cite{chen2020simple}, BYOL \cite{grill2020bootstrap} and SWAV \cite{caron2020unsupervised}, which all assist the learning of input similarity using different contrastive loss function. We do not expect a major enhancement of the  SNN classification performance from any of these contrastive models.

To incorporate the different data structure as inputs to a neural network, a dual-input  HDNN  is then constructed \cite{Kim:2019wns,Huang:2022rne,Flacke:2023eil,Hammad:2022lzo}. The first stream consists of fully-connected layers that process the reconstructed kinematical variables (see below). The second stream consists of two-dimensional convolution layers and pooling layers that process the jet images. The two streams are then concatenated into one flatten layer, then, for better expressivity,  a fully-connected layer is added before the  final output layer. The HDNN model with dual inputs has the advantage to combine the global information captured by the fully-connected layers acting on the kinematical distributions and the local information captured by the kernels in the convolution layers acting on the jet images. To analyze the combined  information, global and local one, exalts the model expressivity in terms of signal and background events, which in turn enhances the overall classification performance.

CNNs are specifically tailored to process grid-like data structures, such as images, where local information is paramount. By using predefined filters, CNNs capture local patterns effectively. However, this design inherently carries significant inductive biases \cite{Biases}. As the constructed jet images are sparse, inductive biases confuse the model, which ends up with lower classification performance. The challenge here stems from the model inability to adequately process sparse, non-grid-like data, which is a significant limitation for CNNs. To address these issues, we propose the use of GNNs instead. Unlike CNNs, GNNs can process input data that naturally form a graph structure, with entities represented as nodes and relationships as edges \cite{Qu:2019gqs}. This makes GNNs adept at handling sparse and/or irregular data. In the case of jet physics, for instance, the four-momenta of the final state particles can be seen as graph nodes, while the graph edges can be weighted by the angular distance between the particles. GNNs have an inherent ability to handle both local and global information in the data. They propagate information across the graph, allowing each node to be influenced by its neighbor information and iteratively capture long-range dependencies. This makes GNNs better performing so as to overcome the limitations of CNNs in this context.

As a generic set-up for all the proposed DNNs, we require all models to have an output layer with two neurons and a softmax function. The loss function is  the categorical cross entropy defined as 
\begin{equation}
    \text{Loss}  = -\sum_i Y_i\log(\hat{Y_i})\,,
\end{equation}
with $i=1,0$ for signal and background classes, respectively, and $Y_i,\hat{Y_i}$ are the true and predicted labels from each class. The dimension of the final output probability, $\hat{Y}$,  is $1\times 2$, $(\mathcal{P}_{sig},\mathcal{P}_{bkg})$, with $\mathcal{P}$ ranging between $[0,1]$. If $\mathcal{P}_{sig} > 0.5\ (\mathcal{P}_{bkg} < 0.5)$, the corresponding event is classified as most likely being a signal event and if $\mathcal{P}_{sig} < 0.5 \ (\mathcal{P}_{bkg} > 0.5)$ the corresponding event is classified as most likely being a background event. An AdamW optimizer \cite{adamw} is used to optimize the minimization of the loss function with learning rate $10^{-3}$, weight decay  $4\times 10^{-3}$ and exponential decay rate $0.9$. The size of the input data is $200,000$ in all models, divided into $70\%$ for training and $30\%$ for testing the model accuracy. The DNNs are trained and tested on equal size data sets for signal and background events. 

it's worth noting that we haven't fine-tuned the proposed networks, given the substantial computational resources that would be necessary. Indeed, conducting a grid search of the hyper-parameters could potentially improve the accuracy of the classification results we've reported.

\subsection{MLP}
\label{sec:MLP}
A MLP is the basic type of a feed-forward DNN which consists of fully-connected hidden layers of different length. Given the nature of the fully-connected layers, a MLP  is designed to learn  the global information in the reconstructed kinematical distributions.  This can be achieved by firing specific neurons in each hidden layer corresponding to the signal or background distributions. After training,  a MLP exhibits specific structures of the fired neurons in case of signal or background events. We point out that the full connection of the MLP hidden layers enables the model to propagate all event information among all hidden layers and thus its ability to learn global information about the event is increased \footnote{Obviously, a MLP cannot be used for jet image analysis as the nature of the fully-connected layers makes the model depend on the spatial position of the energy deposits into the image. In contrast, a CNN with local weight sharing among its kernels makes the model independent of such a spatial position.}. 

\begin{figure}[h!]
\includegraphics[width=\textwidth]{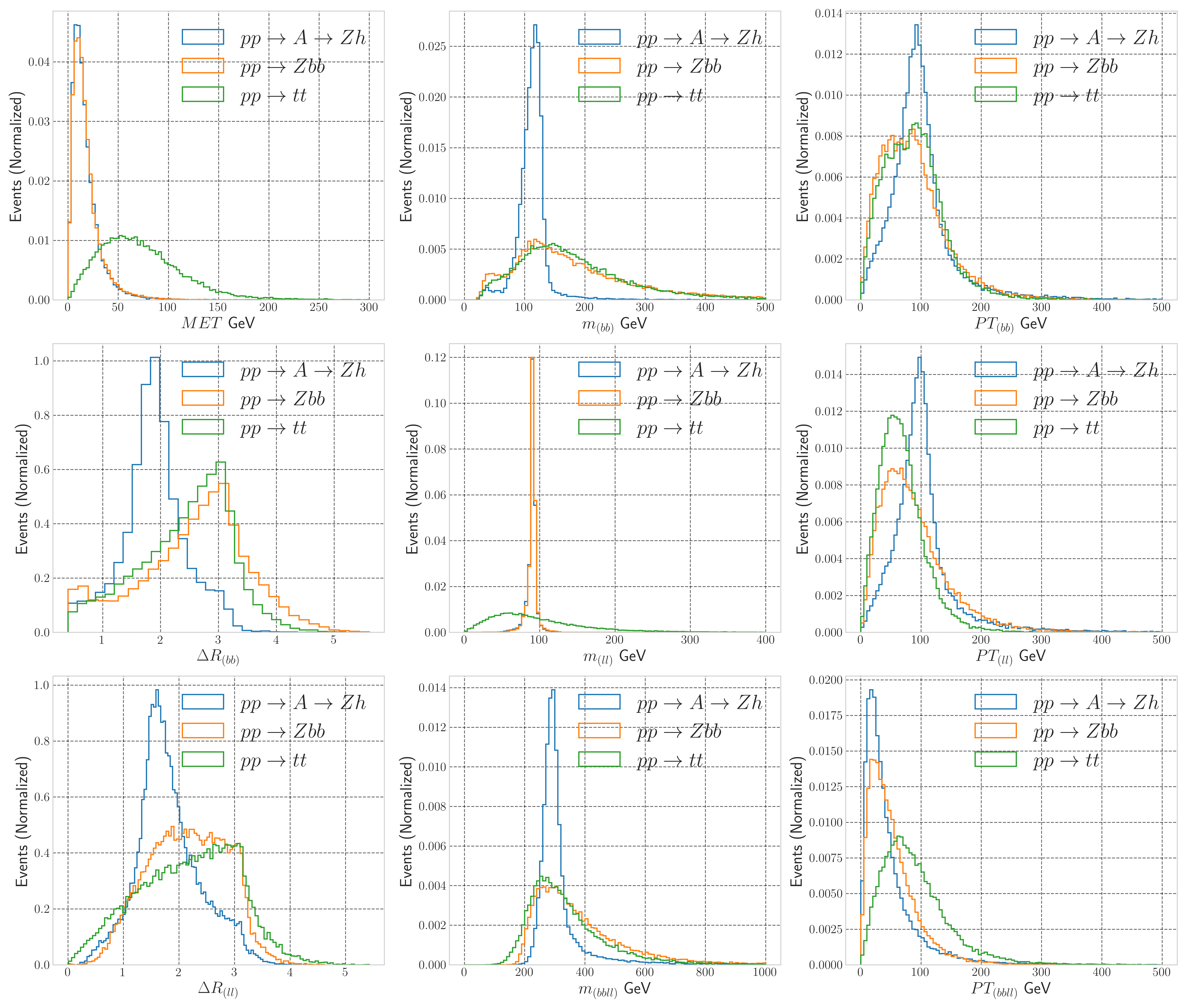}
\caption{Kinematical distributions for signal (BP with $m_A=300$ GeV) and background events superimposed and normalized to 1 before applying the pre-selection cuts. The color codes hold for all distributions as follows: signal (blue), $pp\to t\bar{t}$ (green) and $pp\to Z b\bar{b}$ (orange).}
\label{fig:kinematic}
\end{figure}

For optimal classification performance, we select distributions with high discrimination power between signal and background events.  To select the highly ranked distributions we follow a  Sequential Backward Selection (SBS) feature \cite{SFS} by first constructing all possible kinematical distributions and `greedily'\footnote{That is, by using a greedy algorithm that  follows the problem-solving heuristic of making the locally optimal choice at each stage.}  removing one feature after another, in order to find the one  that maximizes a cross-validated score when an estimator is trained on this single feature. The feature selection method indicates highly ranked nine kinematical distributions as shown in Fig.  \ref{fig:kinematic}, e.g.,  for the signal BP with $m_A=300$ GeV. Although the kinematical distributions herein are for a specific signal point, we found that other signal BPs  have similar discriminative power. The selected kinematical  distributions can be chosen as follows.

\begin{itemize}
\item  {$MET$}: Missing Transverse Energy ($MET$), defined as ${MET}  =|-\sum_{v_i}\vec{pT}(v_i)|$, which is the sum of the transverse momenta of the visible particles.  This is very similar for the signal and $pp\to Zb\bar b$, with $pp\to t\bar t$ set apart.

\item {$m_{(bb)}$:} Invariant mass of the $b$-jet pair. Signal events show a narrow peak around the rest mass of the SM-like Higgs boson while background events show a broader peak as they are initially produced from QCD radiation in the case of $pp\to Zb\bar b$ and from top decays in the case of $pp\to t\bar t$. 

\item {$P{T_{(bb)}}$:} Transverse momentum of the $b$-jet pair
reproducing $m_h$, which exemplifies the Higgs boson boost in  signal events (and is different for other BPs).  This distribution has a strong degree of similarity with the background ones.

\item {$\Delta R_{(bb)}$:} Angular distance separation between the two $b$-jets reconstructing the Higgs boson, with $\Delta R_{(b{b})} = \sqrt{(\Delta\eta_{(b{b})})^2+(\Delta\phi_{(b{b})})^2}$. For the $pp\to Zb\bar b$ background, the two $b$-jets recoil against the associated $Z$  when the latter is produced near its mass shell, thus they have a small boost factor and fly back-to-back with angular distance around $\pi$. A similar behavior
also applies to the $b$-jets emerging from top-(anti)quark  leptonic decays, for which {$\Delta R_{(b{b})}$} peaks again around $\pi$. Signal events show instead a narrow peak around 1.6 (for a  heavier $A$, e.g., $m_A=350$ GeV, the $b$-jets receive extra an  boost and $\Delta R_{(b{b})}$ peaks around 1. 

\item {$m_{(ll)}$:} Invariant mass of the lepton pair. Reconstructed events from the signal and $pp\to Zb\bar b$ processes offer a tight reconstruction of the $Z$ boson mass by  showing a narrow peak around $m_Z$ while,  for  $pp\to t\bar t$ events, the final state leptons emerge from a pair of $W^\pm$ boson decays, thereby missing such a distinctive feature. (In the case of signal events from the BP with $m_A=200$ GeV, the $Z^*$ boson is produced off-shell and thus the invariant mass peak for the signal is very similar to that of $pp\to t\bar t$.)

\item {$P{T_{(ll)}}$:} Transverse momentum of the two  leptons 
 reproducing $m_Z$ which exemplifies the $Z^{(*)}$ boson boost in  signal events (again, the latter and $pp\to t\bar{t}$ events have similar distributions which are in turn different from that of the background process $pp\to Zb\bar  b$). 

\item {$\Delta R_{(l{l})}$:} Angular distance separation between the two leptons reconstructing the $Z^{(*)}$, which exhibits a similar behavior as the angular separation between the two $b$-jets in all cases. 

\item {$m_{(bbll)}$:} Invariant mass of the $b$-jet and lepton pairs which reconstruct the masses of the $h$ and $Z^{(*)}$ boson,
respectively, in turn reconstructing the mass of the $A$ state.
(For the BP with $m_A = 200$ GeV, the reconstructed $A$ mass peak is broader as the final leptons from the off-shell $Z^*$ boson decay are soft and can be missed.)  There is a clear difference here between signal and background events.

\item {$pT_{(bbll)}$:} Transverse momentum of the $b$-jet and lepton pairs which reconstruct the masses of the $h$ and $Z^{(*)}$ boson, respectively, in turn exemplifying the $A$ state boost in  signal events, which overlap significantly with that of background ones.

\end{itemize}

\begin{figure}[h!]
\centering
    \includegraphics[width=9cm, height=4cm]{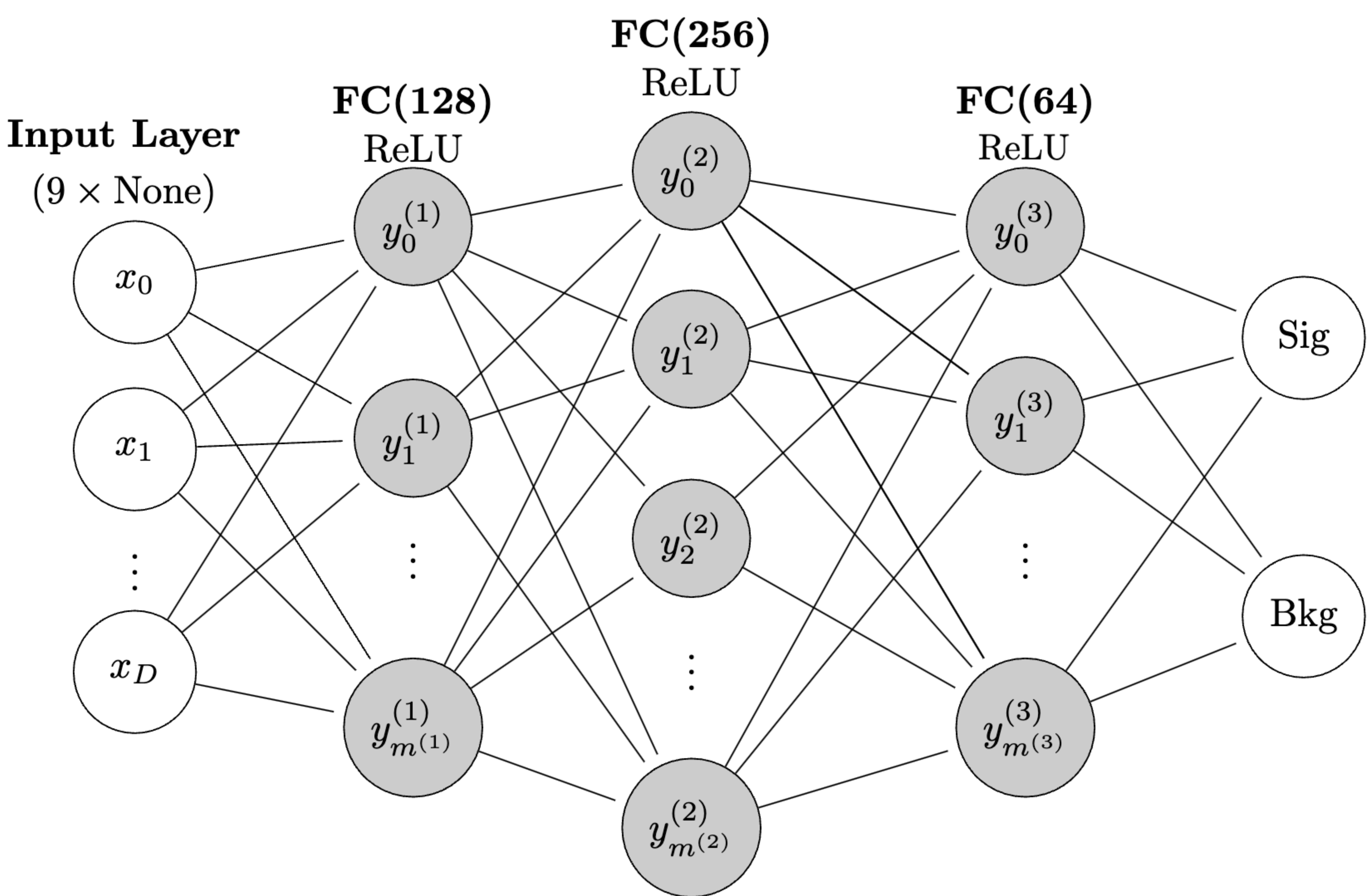}
    \caption{A schematic architecture for the used MLP model. Here, we visualize a Fully-Connected (FC) NN layer. }
\label{fig:MLP_net}
\end{figure}

After reconstruction of the kinematic distributions we stack all backgrounds and signal events separately such that each data set has dimensions $d_{\text{distribution}} = (9,N)$  with $N$ being the total number of events. As a supervised classification problem,  we assign a numeric label $Y=1$ to the signal events  and $Y=0$ to (the whole of) the background events. Having the input distributions and their labels adjusted, we use MLP with an input layer of the same dimension as the inputs. Fig. \ref{fig:MLP_net} shows a schematic architecture of the used MLP model which consists of three pairs of fully-connected hidden layers with Rectified Linear Unit (ReLU) activation function and an output layer with two neurons and softmax activation function. The number of neurons in the first pair is $256$, in the second  is $128$ and in the third is 64. To avoid over-training, each hidden layer pair is followed by a dropout layer. 
During the training process the model tries to minimize the difference between its predictions and the assigned labels. To measure  the model ability to generalize to new unseen data, we test the model accuracy to unseen test set.  

Fig. \ref{fig:1_3} shows the MLP results from the test sample for the signal BP with $m_A=300$ GeV. To quantify the classification power of the model, the left plot shows the  Receiver Operator Characteristic (ROC) curve
The middle plot shows the output score of the model for the signal events (blue) and background events (orange). The right plot shows the CM when a symmetric threshold value at $0.5$ is used on the  model output.
\subsection{CNN}
\label{sec:CNN}
\begin{figure}[h!]
\label{fig:jet_images}
\includegraphics[width=\textwidth]{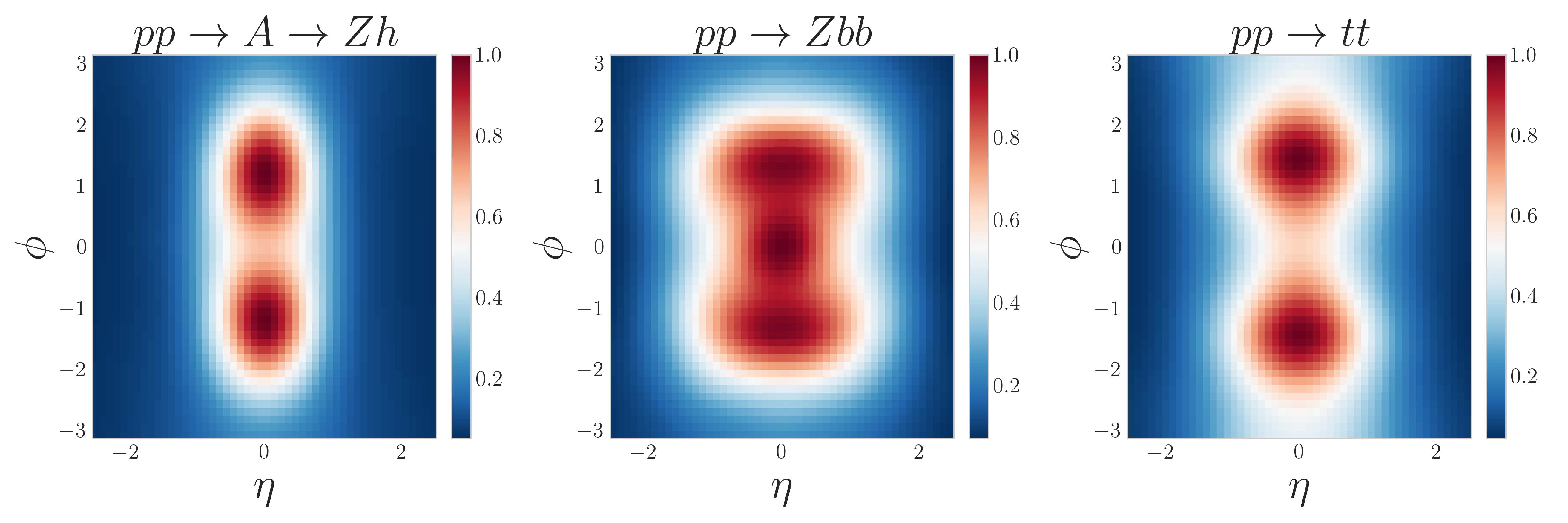}
\caption{Normalized $pT$ distribution for accumulated $50,000$ events after pre-processing steps for signal  (BP with $m_A=300$ GeV) and backgrounds events.}
\end{figure}

As mentioned above,  the global structure of the color flow can be seen at the LHC  as a color string from the soft hadrons that stretch between the two colored connected jets. The different color structure for different processes originates from the color nature of the parent particle, which can provide the event with an observable to aid the search for new physics. The two $b$-quarks from the decay of the Higgs boson form a color dipole whose radiation pattern is contained primarily within a pair of cones around the two $b$-quarks, with a tendency for more radiation to occur in the region between the two. In contrast, the two $b$-quarks in $pp\to Zb\bar b$ and $pp\to t\bar t$ events come from colored particles and  are thus not directly connected, forming two isolated cones with less radiation in the region between the two $b$-quarks that in the signal. To effectively identify the different radiation patterns,  we construct the jet images as a squared array in the ($\eta, \phi$) plane with each pixel given by the total hadron $pT$ deposited in the associated region of the calorimeter. In Fig. \ref{fig:jet_images} we show  normalized $pT$ distributions for $50,000$ events for signal BP with $m_A=300$ GeV (left), $pp\to Zbb$ (middle) and $pp\tt$ (right). 
To ensure that the CNN is not learning space-time symmetries and can be generalized to new unseen data at different  locations, the jet images are pre-processed as follow:

\begin{enumerate}
\item \underbar{Image cleansing}  ~Images are constructed only from hadrons which have track information while at the same time we  remove leptons (and photons).

\item \underbar{Pixelization}  ~The region in the $(\eta, \phi)$ plane is discretized  into a $50\times 50$ grid with each pixel weighted by the sum of the transverse momentum in it.

\item \underbar{Centering} ~We center all particles in an image by shifting $( \frac{(\eta_{b} +\eta_{\bar{b}})}{2} , \frac{(\phi_{b}+\phi_{\bar{b}})}{2})$ to  $(0, 0)$, which assists the independence of the model classification from the spatial location of the radiated hadrons.

\item \underbar{Momentum smearing} ~Constructed images are mostly sparse, which hinders the classification performance of the CNN model. To reduce the number of the sparse pixels, we smear the transverse momentum using a Gaussian function within 3 Standard Deviations (SDs) or $\sigma$'s \cite{Buss:2022lxw}.

\item \underbar{Normalization}  ~We normalize the pixel intensity by dividing each pixel in an image by the maximum pixel intensity value, which helps the model to converge to the global minimum of the loss function.
\end{enumerate}

\begin{figure}[h!]
\centering
    \includegraphics[width=12cm, height=4cm]{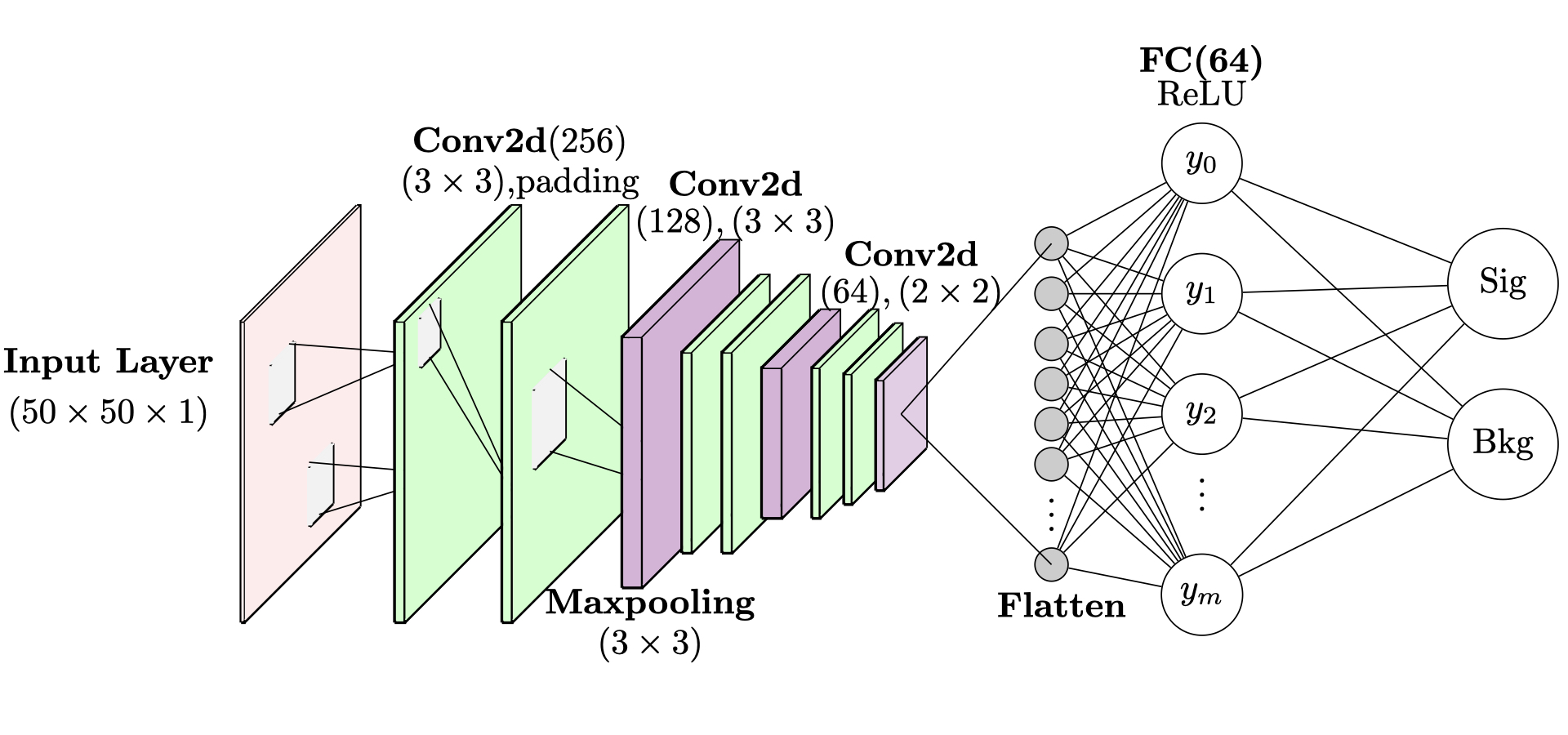}
    \caption{A schematic architecture for the used CNN model. Here, `Conv2d' represents a two-dimensional CNN layer. }
\label{fig:CNN_net}
\end{figure}

After the pre-processing steps, the constructed image has the dimension of  $50\times 50\times 1$. In principle, one can add more information to the images, e.g., leptons, $MET$, etc. properties \cite{Kim:2019wns}. That information can be incorporated into an image by expanding the last dimension of it, i.e., the image depth. Although having more information into an image allows the model to learn more characteristic features of the events, we found that including leptons and $MET$ information to our images does not increase the classification performance very much. Instead, we opted not to include such an information in order to reduce the computational costs. 

To analyze the constructed jet images we adopt a CNN model with the structure depicted in Fig. \ref{fig:CNN_net}. The model consists of  four convolution layers, one fully-connected layer and one output layer. The first and second convolution layers have $256$ kernels with kernel size $3$, ReLU activation function and stride length of $1$. To keep the dimensions of the original input images, we use a padding layer. Third and fourth convolution layers have $128$ kernels with kernel size $3$ and ReLU activation function. Fifth and sixth convolution layers have $64$ kernels with kernel size $2$ and ReLU activation function. After the second and the fourth convolution layers we use max pooling layers with size $2\times 2$. After these, we use a dropout layer with $30\%$ dropout rate. Output from the last convolution layer is flattened and projected to one fully-connected layer and dropout layer with $64$ neurons and ReLU activation function.


The results of the CNN analysis are shown in Fig. \ref{fig:1_3}. The classification performance is here quantified by the reported AUC metric with value $0.86$. We finally notice that the classification  performance of the CNN analysis of the jet images can be enhanced with extra pre-processing steps, e.g., by constructing the Lund plane \cite{Khosa:2021cyk}  or Riemannian mapping \cite{Hammad:2022msl}.  
\subsection{SNN}
\label{sec:siamese}
The SNN was first introduced as an algorithm for handwritten signature verification \cite{siamese_1}. The main power of the SNN is that it maps input features into a latent space such that a simple distance in it, the Euclidean distance, approximates the characteristic features  in the original one. It consists of two identical convolution encoders sharing the same set of weights in order to compare a pair of feature vectors in terms of their similarity or dissimilarity. It realizes a non-linear embedding of the data with the objective of bringing together similar examples and to move apart dissimilar examples.
To measure the similarity or dissimilarity of the input pairs we use the Euclidean distance as a similarity metric learning given by 
\begin{equation}
    D = \sqrt{\sum_i^n \left( x^1_i - x^2_i\right)^2}\,,
\end{equation}
where $x^1$ and $x^2$ are the latent outputs from the two encoders and $n$ is the latent space dimension. More precisely, given a pair of input images,  the two encoders extract the features in each image and map these onto the latent space as vectors $(x^1,x^2)$. The SNN then minimizes the Euclidean distance between $x^1$ and $x^2$ if they belong to the  same class, e.g., the signal or background class, while it maximizes the Euclidean distance between $x^1$ and $x^2$ if they belong to different classes, e.g., signal and background classes. To do so, the SNN has to be trained in two stages. Firstly, the model computes the similarity or dissimilarity by minimizing a modified contrastive loss function as 
\begin{equation}
    \mathcal{L}(y,D) = \alpha(1-y)\ast D^2 + y\ast \left[\text{Max}(\beta -D,0) \right]^2\,,
\end{equation}
with $y,D$ being the true and predicted distance, respectively, while $\alpha,\beta$ are the margin parameters for learning the similarity and dissimilarity, respectively. Both parameters are hyper-parameters to be tuned (in our study we fix both to 1). Also, we adjust the true distance between the negative pair to be 1 and 0 for the positive pair. We would also like to mention that, in the self-supervised contrastive learning, data augmentation is used for learning similarity and dissimilarity  \cite{Dillon:2021gag,Dillon:2023zac,clr}. In this case, the classification performance depends on the impact of the data augmentations \cite{zhang2022}. Moreover, strong augmentations and implicit regularization may cause dimensional collapse of the projected data into the latent space \cite{jing2021}. We stress that, for our SNN with supervised contrastive learning, the mentioned problems no longer exist.

\begin{figure}[h!]
\centering

\includegraphics[width=14cm, height=5.5cm]{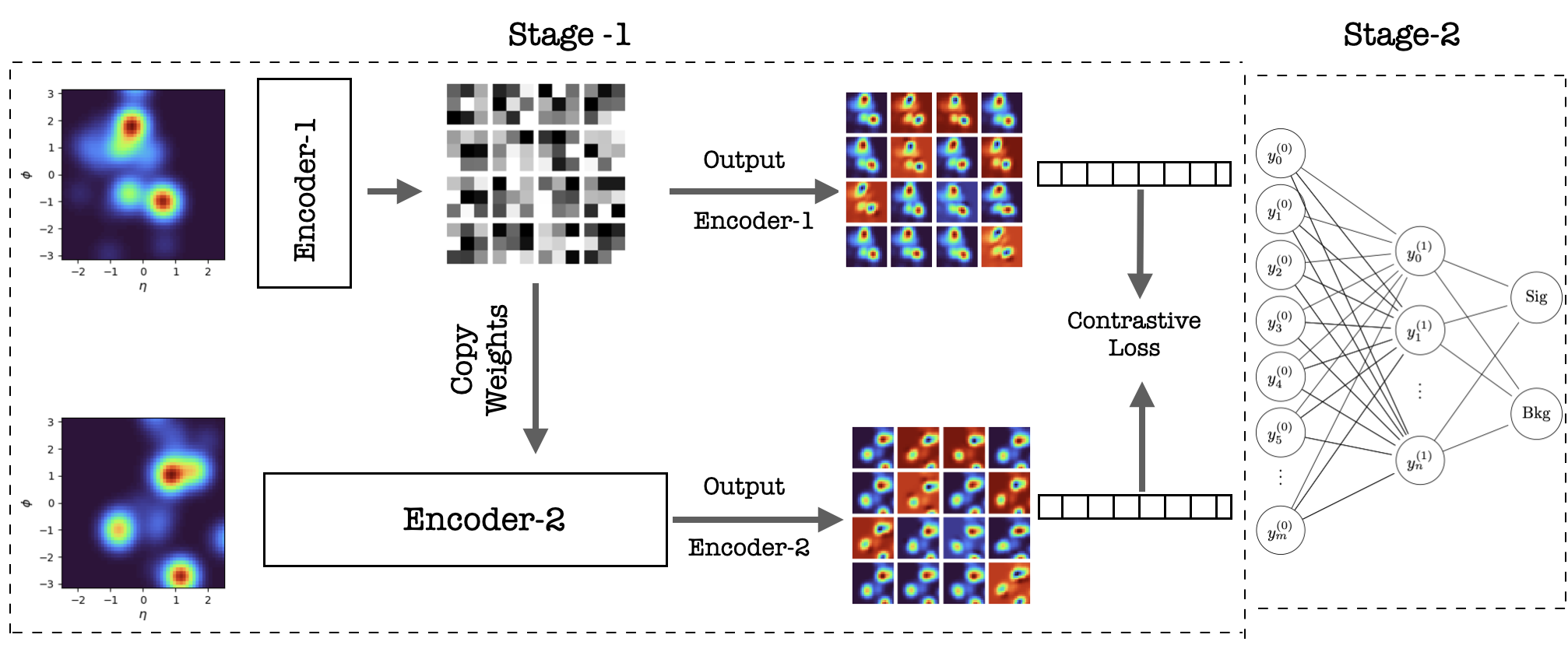}
\caption{A schematic architecture for the used SNN model. Here,  Encoder-1 and Encoder-2 are identical and copy their weights during the first training stage. Input images can be a positive pair, i.e., both images are either signal or background, or a negative pair, i.e., one signal image and one background image.}
\label{fig:siamese_net}
\end{figure}

Once the model is trained to minimize the contrastive loss function, we start the second learning stage by freezing the weights of one of the encoders and add two fully-connected layers and one output layer  with two neurons and softmax function. For the training in the second stage, signal images are labelled with $1$ while background images are labelled with 0 and we use a categorical cross entropy loss function. A schematic architecture of the used SNN model is shown in Fig. \ref{fig:siamese_net}, which consists of two identical convolution  encoders, each of these having the same structure as the discussed CNN without the output layer.   
 
The results of the SNN are shown in Fig. \ref{fig:1_3}, which shows a larger classification performance over the used CNN with $AUC=0.95$. Although the SNN processes the same jet images as the CNN network, it shows an improved classification accuracy over the CNN. This enhancement is, obviously, due to the minimization of the contrastive loss function, which in turn enables the model to learn more information from the jet images. (A detailed discussion about the learned representations by the hidden layers and its impact on the classification performance is presented in section \ref{sec:5}.)

\subsection{HDNN}
\label{sec:mltui_inputs}
To improve the expressivity of the ML model of  signal and background events one can incorporate different information into the discussed models, e.g., by adding the lepton information to the reconstructed images to be analyzed by the CNN or encoding the hadrons information as distributions to be analyzed by the MLP. In both cases one can find a slight improvement in the classification performance of each model individually, as the latter is still able to learn specific types of event information, local or global. Furthermore, concatenating a MLP and CNN into a two stream HDNN model can improve the classification performance \cite{Kim:2019wns,Huang:2022rne,Hammad:2022lzo,Flacke:2023eil}.
\begin{figure}[h!]
\centering
    \includegraphics[width = 12cm, height = 4.8cm]{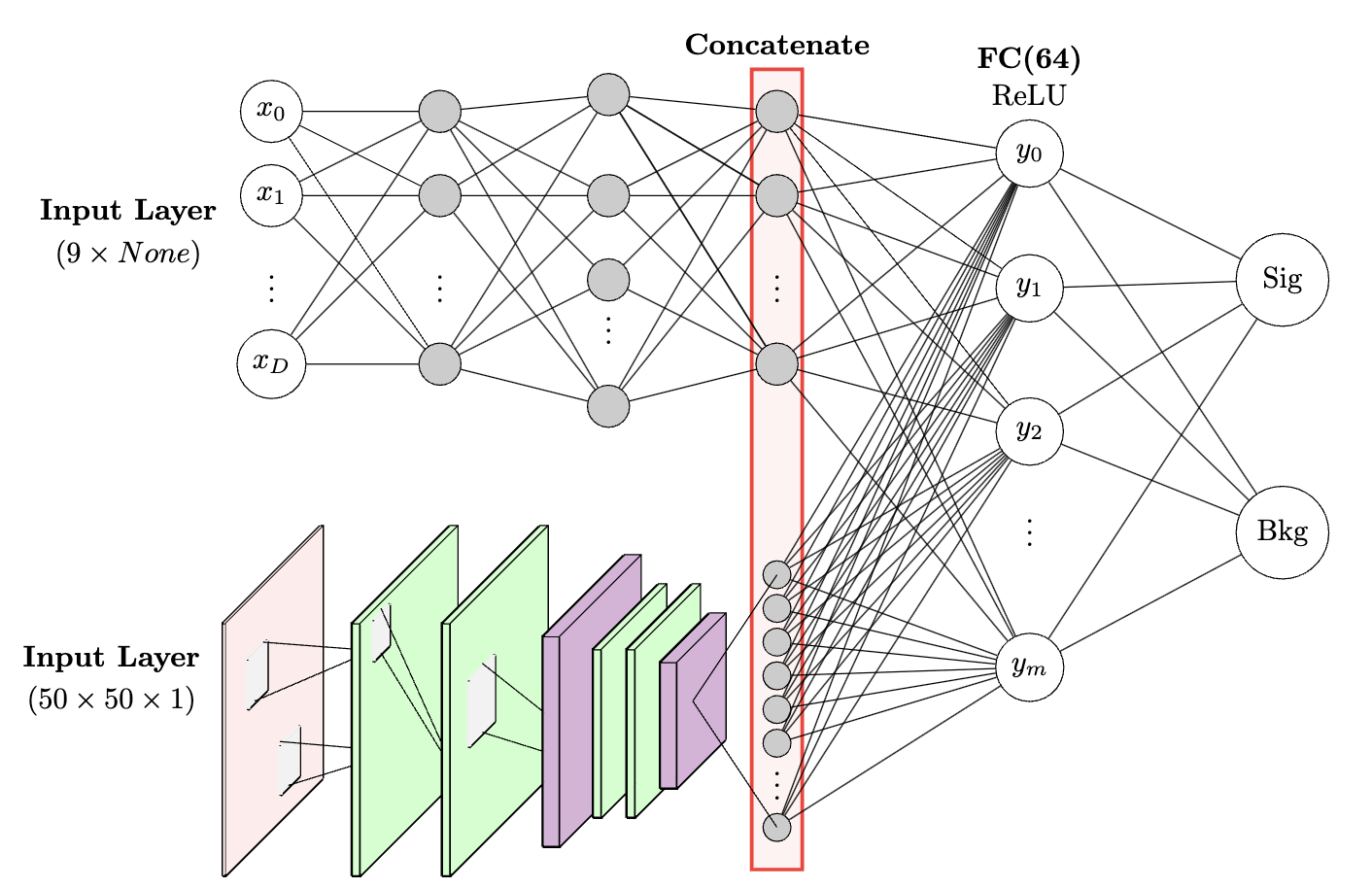}
    \caption{A schematic architecture for the used HDNN model.}
\label{fig:HDNN_net}
\end{figure}
The first stream, which processes the input jet images, consists of convolution, max pooling and drop-out layers plus one flattened layer. The second stream, which processes the kinematic distributions, consists of fully-connected  and drop-out layers. Both streams are then concatenated to one 
fully-connected layer and one output layer with two neurons for predictions, a HDNN. 
The two streams map the high dimensional information onto their own latent space (a lower dimensions space), e.g., the CNN and MLP map the local and global high dimensional information onto lower dimensional space individually. Concatenating the decomposed information in each latent space into one flatten layer expresses  all  characteristic  features of the signal and background events\cite{Chakraborty:2019imr}. A schematic architecture of the HDNN is shown in Fig. \ref{fig:HDNN_net}.

The HDNN is constructed by combining the above  CNN convolution layers and MLP without the output layer. A concatenation layer is used to connect the last layer of the two models. A fully-connected layer with $128$ neurons is added with ReLU activation function and a dropout layer with a $30\%$ dropout rate. The last output layer consists of two neurons and a softmax activation function.

The results of the HDNN analysis are shown in Fig. \ref{fig:1_3}.  The ROC curve shows a large improvement in the HDNN classification performance over the MLP or CNN individually, with $AUC=0.98$. Moreover, incorporating the different data types into a HDNN with two streams enables the model to learn the intrinsic  features of the signal and background events with equal rates as reported by the confusion matrix (right plot). 
\subsection{GNN}
\label{sec:intro}
One way to avoid the sparsity issue of image-based NNs is to utilize a graphical structure consisting of nodes and edges to encode particle information\cite{ExaTrkX:2020nyf,Abdughani:2018wrw,Qasim:2019otl,Abdughani:2020xfo}. GNNs can then be employed to incorporate the topological relationships among the nodes and edges and learn graph-structured data. Each reconstructed final state object is represented by a single node in this approach. This study, in alignment with the methodology described in \cite{Huang:2022rne}, represents each node $i$ in the input layer as a feature vector $\textbf{x} = (I_{l}, I_{b}, m, p{T}, E)$ that collects the properties of the corresponding particle, where, e.g., $m$, $p{T}$ and $E$ denote the invariant mass, transverse momentum and energy of a particle system, respectively. The initial values for $I_{l}$ and $I_{b}$ are set to 0. The hardest lepton and $b$-jet in an event assign a value of 1 to $I_{l}$ and $I_{b}$, respectively, while, the second hardest lepton and $b$-jet in the event assign a value of -1 to $I_{l}$ and $I_{b}$, respectively. The angular correlation between two nodes $i$ and $j$ is represented by an edge vector $e_{i,j}$, which consists of a single component that is defined by the angular distance $\Delta R_{(x_{i}, x_{j})}$ between the particles in nodes $i$ and $j$.

In the course of our comprehensive study, various architectures were tested to identify the optimal model capacity. Subsequent to extensive trials and analysis, our observations revealed that the maximum performance across all tested GNN models was achieved with a configuration of three hidden layers. 

This preference can be rationalized by considering the nature of the graphs involved, typically containing a limited number of nodes, that is, ranging between 4 and 25. Each layer in a GNN, by design, corresponds to the aggregation of information from the neighboring nodes, which is one edge away (1-hop). With small-scale graphs, only a few layers are often sufficient to incorporate the entirety of the graph data. 
Conversely, incorporating an excessive number of layers in a GNN can potentially lead to an undesirable effect known as oversmoothing. This is a circumstance in which the features of all nodes become overly homogeneous, subsequently impairing the performance of the model.

In terms of activation function, we employed the Leaky Rectified Linear Unit (LeakyReLU) following graph convolution layers. This was then followed by the utilization of a max pooling layer to aggregate the node embeddings, subsequently applying a final linear layer.
For the optimization process, a learning rate of 6.4 $\times$ 10$^{-6}$ and a weight decay parameter of 1 $\times$ 10$^{-6}$ were employed. These values were chosen to ensure efficient learning without compromising the stability of the model.
All of the developed models in our study were constructed utilizing the `PyTorch Geometric' framework \cite{fey2019fast}, a powerful and efficient library designed to facilitate the implementation of graph-based DL models.


\subsubsection{GCN}
GCNs have gained significant attention in recent years due to their ability to learn representations of graph-structured data that are invariant with the input graph size and topology \cite{kipf2016semi}. GCNs have been successfully applied to various domains, including social network analysis, molecular biology, recommendation (or recommender) systems and natural language processing. The goal of a GCN is to learn a function that maps the input features to a new representation, capturing the relationships among the vertices in the graph. 

The core idea behind GCNs is to generalize the convolution operation from regular grids to irregular graphs. A graph convolution operation can be thought of as a local averaging of features from neighboring vertices, which captures both the local structure of the graph and the features associated with each vertex.

Given an input graph $G = (V, E)$, the graph convolution operation is defined as
\begin{equation*}
H^{(l+1)} = \sigma \left( \hat{D}^{-\frac{1}{2}} \hat{A} \hat{D}^{-\frac{1}{2}} H^{(l)} W^{(l)} \right),
\end{equation*}
where $H^{(l)} \in \mathbb{R}^{N \times F_l}$ is the feature matrix at layer $l$, with $N$ being the number of vertices in the graph, $F_l$ the dimension of the feature space at layer $l$ and $W^{(l)} \in \mathbb{R}^{F_l \times F_{l+1}}$ the learnable weight matrix at layer $l$. Furthermore, $\sigma(\cdot)$ denotes the activation function.

The matrix $\hat{A} \in \mathbb{R}^{N \times N}$ is the adjacency matrix of the input graph with added self-connections, defined as $\hat{A} = A + I_N$, where $A$ is the adjacency matrix of $G$ and $I_N$ is the identity matrix of size $N$. The matrix $\hat{D} \in \mathbb{R}^{N \times N}$ is a diagonal matrix with $\hat{D}_{i} = \sum_j \hat{A}_{ij}$, representing the degree of vertex $i$ in the graph with added self-connections.

The graph convolution operation can be interpreted as a message-passing mechanism, where each vertex aggregates information from its neighbors and updates its features according to the learned weights. This process is repeated for a number of layers, allowing the model to capture higher-order relationships between vertices in the graph.

\begin{figure}[tbh!]
\centering
    \includegraphics[scale=0.6]{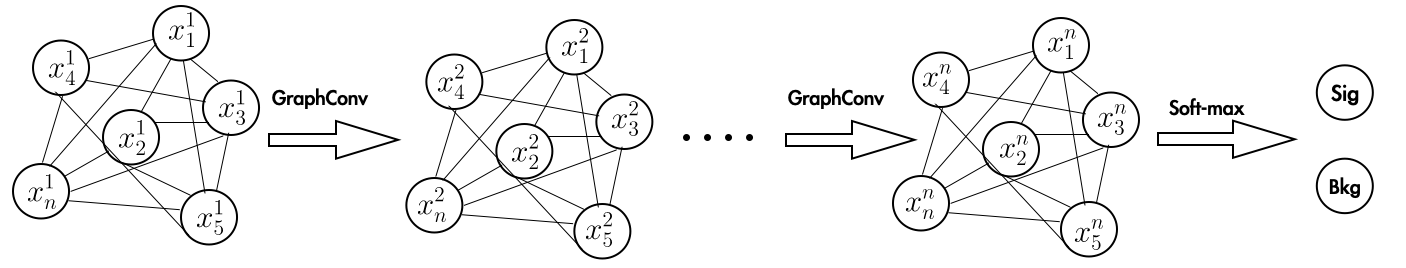}
    \caption{A schematic architecture for the used GCN model.}
\label{fig:GCN_net}
\end{figure}

The results from our GCN analysis are presented in Fig. \ref{fig:1_3}. The leftmost plot displays the ROC curve for the trained GCN model, showcasing an AUC value of 0.84. The middle plot demonstrates the model scores for both signal and background, providing a visual representation of the model's predictive distributions. On the rightmost plot, the CM is displayed where the signal and background diagonal entries are 0.83 and 0.76, respectively.


{Despite the demonstrated effectiveness of the GCN model, it isn't without its constraints. These limitations have encouraged the exploration and establishment of multiple variants and extensions. One major constraint of the original GCN model is its incapability to handle inductive learning tasks. This inability to generalize to unseen graph structures or vertices is due to the GCN's reliance on the explicit adjacency matrix of the input graph, which makes adapting the model to new data a challenge.

Additionally, the model lacks flexibility when capturing a wide array of graph structures and applying filters of different spatial sizes, primarily because the conventional GCN model utilizes a fixed neighborhood aggregation scheme. It's also worth noting that GCNs treat all neighboring vertices with equal importance during the feature aggregation phase, which could be sub-optimal when some neighbors provide more significant information than others.

These shortcomings led to the development of diverse GCN variants, including GraphSAGE, Graph Attention Networks (GAT), and Dynamic Graph Convolutional Neural Networks (DGCNN). These adaptations address the challenges by incorporating inductive learning capabilities, applying spectral techniques, and/or introducing attention mechanisms, thereby extending the usability and enhancing the performance of graph-based deep learning models.
}
\subsubsection{DGCNN}
Unlike GCNs, which often assume that graphs are static in nature and hence fail to capture the dynamics of edges, in a  DGCNN \cite{Wang2019}, the EdgeConv operation serves as the fundamental building block. It considers edges as basic units of information propagation instead of nodes, which is especially beneficial in capturing local geometric structures and dealing with unordered point sets. DGCNNs extend the idea of CNNs to graphs, where each layer of the network operates on the nodes and edges of the graph. What makes DGCNNs unique is their ability to learn the importance of edges dynamically during training. Instead of relying on pre-defined edge weights ($\Delta R$ in our case), a DGCNN learns to assign weights to the edges of the graph based on their importance for the task at hand. This allows the network to adapt to different graphs and tasks more effectively.

For an edge $e_{ij}$, the EdgeConv operation is:
\begin{equation*}
\mathbf{h}_{ij} = \Phi(\mathbf{v}_i, \mathbf{v}_j - \mathbf{v}_i).
\end{equation*}
In this equation, $\mathbf{v}i$ is the feature vector of the node $i$, $\Phi$ is a shared MLP applied to the concatenation of the feature vector of node $i$ and the difference between the feature vectors of nodes $j$ and $i$.
The new feature of node $i$ is then computed by aggregating the transformed features of all neighboring nodes:
\begin{equation*}
\mathbf{v}_i' = \rho(\{\mathbf{h}_{ij} | \forall j \in N(i)\}),
\end{equation*}
where $N(i)$ denotes the neighborhood of node $i$ and $\rho$ is a symmetric function, such as max pooling or average pooling.
The edge features are recomputed in every layer, allowing the graph structure to dynamically evolve based on the learned node features. This dynamic nature is a key advantage of the EdgeConv operation in DGCNNs.


The results of the DGCNN analysis is shown in Fig. \ref{fig:1_3}. This showcases an enhancement from the previous GCN model, with a reported AUC metric value of 0.87, signifying improved classification performance. The dynamic nature of the DGCNN is particularly beneficial in capturing local geometric structures and managing unordered point sets. Furthermore, DGCNNs extend the concept of CNNs to graphs, operating on both nodes and edges through each network layer. 
\subsubsection{GraphSAGE}
GraphSAGE is an inductive learning framework for graph-structured data that allows GCN to generalize to unseen graph structures or vertices \cite{hamilton2017inductive}. Unlike traditional GCNs, which are transductive and rely on the explicit adjacency matrix of the entire input graph, a GraphSAGE structure learns to generate embeddings for individual vertices by sampling and aggregating features from their local neighborhoods.

The core idea behind GraphSAGE is to learn a function that generates vertex embeddings by aggregating features from a fixed-size local neighborhood, irrespective of the graph size or structure. To achieve this, GraphSAGE employs a two-step procedure: sampling and aggregation.

\begin{figure}[tbh!]
\centering
    \includegraphics[scale=0.55]{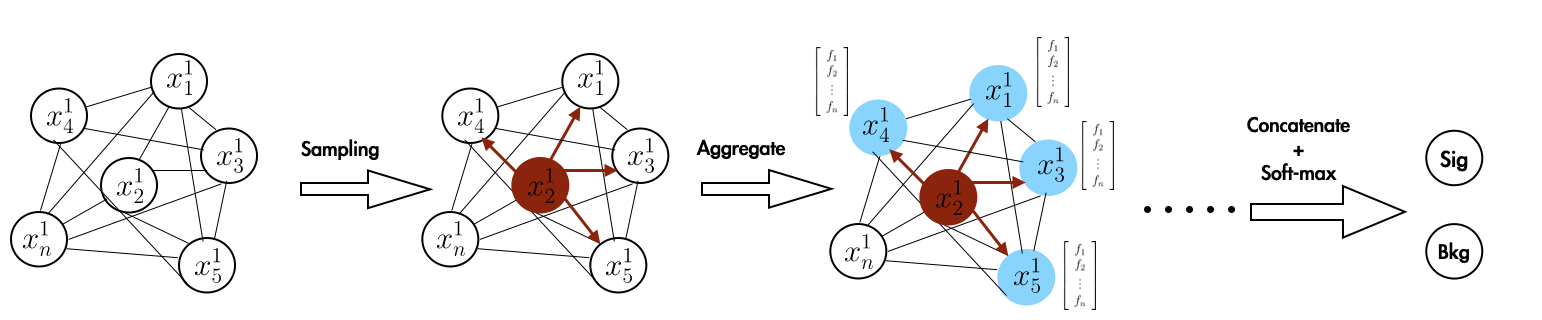}
    \caption{A schematic architecture for the used GraphSAGE model (for the node $x_2$). Here, dots indicates repeated  sampling and aggregation for all graph nodes.}
\label{fig:GCN_net}
\end{figure}

\underbar{Sampling} ~For each vertex $v$ in the graph, GraphSAGE samples a fixed-size set of neighbors at different search depths. The sampling procedure is carried out for $K$ iterations, where $K$ is the number of layers in the model. In the $k$-th iteration, a fixed-size set of $S_k$ neighbors is sampled uniformly at random from the $k$-hop neighborhood of $v$.

\underbar{Aggregation} ~After sampling the neighbors, GraphSAGE aggregates the features from the sampled neighborhood to generate the embeddings for each vertex. The aggregation function can be any differentiable and permutation-invariant function, such as the element-wise mean, Long Short-Term Memory (LSTM) or max pooling. The aggregation process is carried out in a hierarchical manner, starting from the outermost layer and moving towards the target vertex.

Given a vertex $v$, let $\mathcal{N}_k(v)$ denote the set of $k$-hop neighbors of $v$. The feature aggregation process in GraphSAGE can be formally defined as follows:
\begin{equation}
\mathbf{f}_v^{(k)} = \text{AGGREGATE}_k \left(\mathbf{f}_u^{(k-1)} : u \in \mathcal{N}_k(v)\right),
\end{equation}
where $\mathbf{f}_v^{(k)}$ is the feature vector of vertex $v$ at the $k$-th layer and $\text{AGGREGATE}_k(\cdot)$ is the aggregation function at layer $k$.
After aggregating the features from all layers, the final embedding for vertex $v$ is computed by concatenating the original feature vector $\mathbf{f}_v^{(0)}$ and the aggregated feature vector from the last layer $\mathbf{f}_v^{(K)}$:
\begin{equation}
\mathbf{f}_v' = \text{CONCAT} \left( \mathbf{f}_v^{(0)}, \mathbf{f}_v^{(K)} \right).
\end{equation}


The outcomes from the GraphSAGE analysis are displayed in Fig. \ref{fig:1_3}. It reveals a classification performance, measured by the ROC AUC of 0.86. While GraphSAGE is predominantly designed for larger graphs, and our dataset comprises mainly smaller graphs. Despite this, there's still an observed improvement in comparison to the GCN model.

The observed improvement with GraphSAGE on smaller graphs could possibly be attributed to its distinctive feature aggregation method. Unlike GCN, which heavily relies on the graph's global structure, GraphSAGE employs a more flexible, inductive approach, aggregating features from the local neighborhood of each node. As a result, even with smaller graphs, GraphSAGE can extract meaningful, context-rich features leading to enhanced performance. Also, GraphSAGE's sample-based training method helps to capture and generalize even subtle patterns present in smaller graphs.
\subsubsection{GAT}
A GAT is a variant of the GCN that incorporate the attention mechanism to adaptively weigh the importance of neighboring vertices during the feature aggregation step \cite{velivckovic2017graph}. By assigning different weights to neighbors based on their relative importance, GATs are able to learn more expressive and flexible graph representations compared to standard GCNs.

The attention mechanism in GAT is designed to compute a pair-wise attention coefficient between any two connected vertices, which is used to weigh the contribution of neighboring features during the aggregation step. Formally, the attention coefficients for a vertex $i$ and its neighbor $j$ can be defined as:
\begin{equation}
e_{ij} = \text{LeakyReLU} \left( \mathbf{a}^\top \left[ \mathbf{W} \mathbf{h}_i \oplus \mathbf{W} \mathbf{h}_j \right] \right),
\end{equation}
where $\mathbf{h}_i$ and $\mathbf{h}_j$ are the feature vectors of vertices $i$ and $j$, respectively, $\mathbf{W} \in \mathbb{R}^{F' \times F}$ is a shared weight matrix that projects the input features onto a higher-dimensional space, $\oplus$ denotes concatenation, $\mathbf{a} \in \mathbb{R}^{2F'}$ is a learnable attention vector and LeakyReLU is used as the activation function.
To ensure that the attention coefficients are invariant to the order of vertices, the attention mechanism is made symmetric by considering the concatenation of the transformed feature vectors for both vertices. The attention coefficients are then normalized using the softmax function to obtain the final attention weights:
\begin{equation}
\alpha_{ij} = \text{softmax}j(e_{ij}) = \frac{\exp(e_{ij})}{\sum_{k \in \mathcal{N}_i} \exp(e_{ik})},
\end{equation}
where $\mathcal{N}_i$ is the set of neighboring vertices of vertex $i$.

\begin{figure}[tbh!]
\centering
    \includegraphics[width=14cm,height=3.4cm]{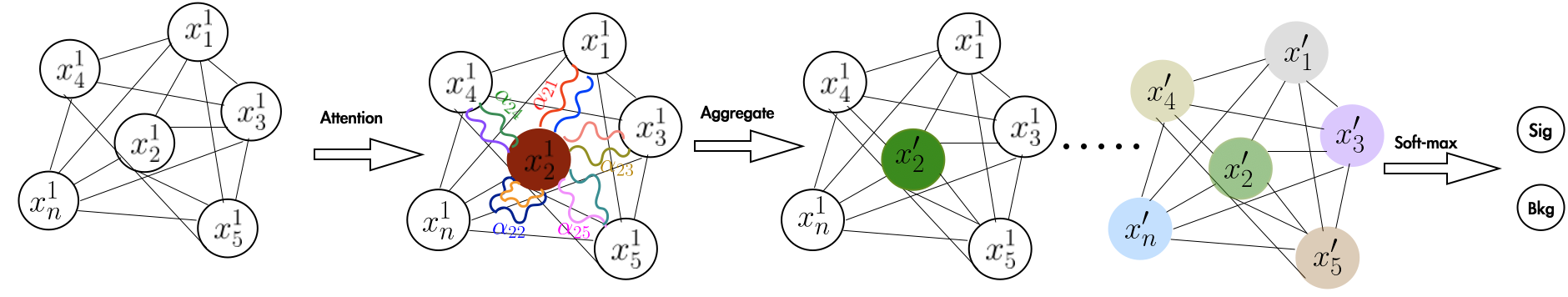}
    \caption{A  schematic architecture for the used GAT model (for the node $x_2$). Here, wavy lines illustrate the self and neighbours attention corresponding to the node while different colors  denote independent attention computations (attention heads). The aggregated features from each head are averaged to obtain $x^\prime_i$. Also, $\alpha_{ij}$ are the attention weights.}
\label{fig:GAT_net}
\end{figure}

With the attention weights computed, the next step in GAT is to aggregate the features from neighboring vertices. The feature aggregation can be expressed as a linear combination of the transformed features of the neighbors, weighted by the attention coefficients:
\begin{equation}
\mathbf{h}i' = \sigma \left( \sum{j \in \mathcal{N}_i} \alpha_{ij} \mathbf{W} \mathbf{h}_j \right),
\end{equation}
where $\sigma(\cdot)$ is the activation function.


The analysis of the GAT is depicted in Fig. \ref{fig:1_3}, with the classification performance, as expressed through AUC metric, returning a value of 0.85. In our training process, we employed four attention heads. This setup allows the model to capture different types of relationships and multi-dimensional information from neighboring nodes, enhancing its representation learning capability. GAT's improvement over GCN could be attributed to its capability to weigh the importance of neighboring nodes differently when aggregating information. 
\par
{In the context of complex graphs, such as those representing relationships amingst final state particles in detectors, GCN, GAT, DGCNN and GraphSAGE each present unique advantages and limitations. GCNs are foundational but may struggle with complex graphs and are susceptible to over-fitting due to small graph sizes. GATs offer attention mechanisms for nuanced feature aggregation but incur computational overheads and may not capture complex edge dynamics effectively. DGCNNs employ EdgeConv operations to efficiently capture both node and edge features, making them a balanced choice in terms of computational efficiency and complexity. GraphSAGE, instead, employs a flexible, inductive approach to feature aggregation from local neighborhoods, making it particularly effective for smaller graphs: its sample-based training method allows it to capture and generalize even subtle patterns, thus leading to enhanced performance. Among these, though, DGCNNs stand out for their ability to capture complex relationships efficiently, explaining their superior performance over GATs in specialized scenarios.} 
\section{Similarity of DNN hidden layers representations}
\label{sec:5}
DL models are treated as computational tools which predictions are very hard to explain according to the learned information in the hidden layers. Recently, there have been proposed interesting methods trying  to explain the predictions of DL models \cite{Barda,rudin2019stop}. They assume that the contribution of a feature can be determined by measuring how the prediction score changes when the feature is altered.  Although the proposed method can explain the change in the prediction score among different types of input features, it does not give a clear insight about what is the learned representation for each hidden layer.  The challenge in analyzing the hidden layers representations of NNs is that features are distributed across a large number of neurons. Moreover, hidden layers do not have fixed size of neurons. Linear CKA \cite{cka} addresses these challenges, enabling quantitative comparisons of representations within and across networks.

To compute the similarity of the hidden layers representations, CKA takes as an input  the hidden layers activation matrices as  $X\in\mathbb{R} ^{d\times P_1}$ and $Y\in\mathbb{R} ^{d\times P_2}$ with $P_1$ and $P_2$ being the neurons in the different hidden layers evaluated on the same input set with size equals to  $d$. The CKA similarity is defined as
\begin{equation}
    \text{CKA}(M,N) = \frac{\text{HSIC}(M,N)}{\sqrt{\text{HSIC}(M,M)\text{HSIC}(N,N)}} \,, 
\end{equation}
 where $M=XX^\top$ and $N=YY^\top$\footnote{As we compute the linear CKA then $M$ is simply $XX^\top$ while, for kernel CKA, $M$ can be computed as $\Phi(X,X)$ with $\Phi$ being the used kernel function.} denote the Gram matrices for the two hidden layers. The main advantage of having a Gram matrix is that we can compute the similarity of hidden layers representations with different number of neurons as the Gram matrix always has the dimension of $d\times d$. Moreover, one can compute the CKA similarity for hidden layers from different DNNs.

\begin{figure}[h!]
    \includegraphics[scale=0.29]{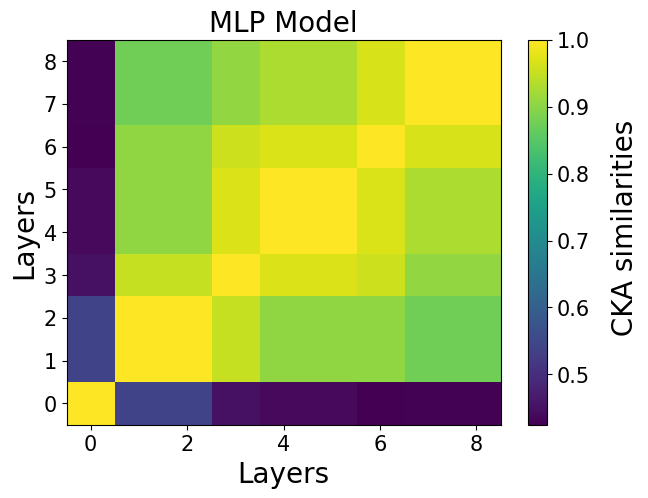}~~
    \includegraphics[scale=0.29]{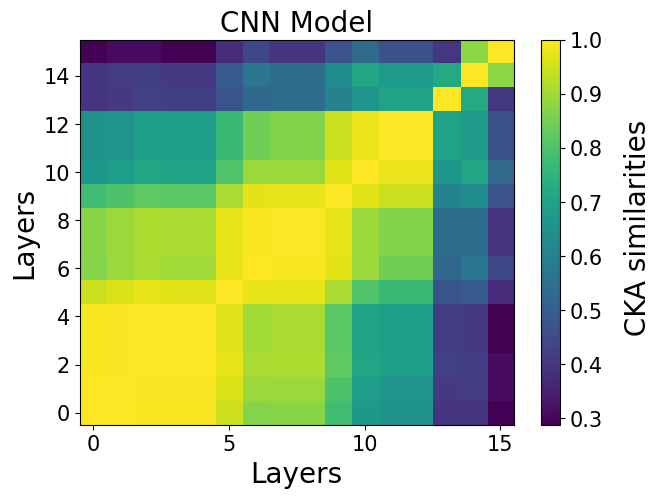}~~
    \includegraphics[scale=0.29]{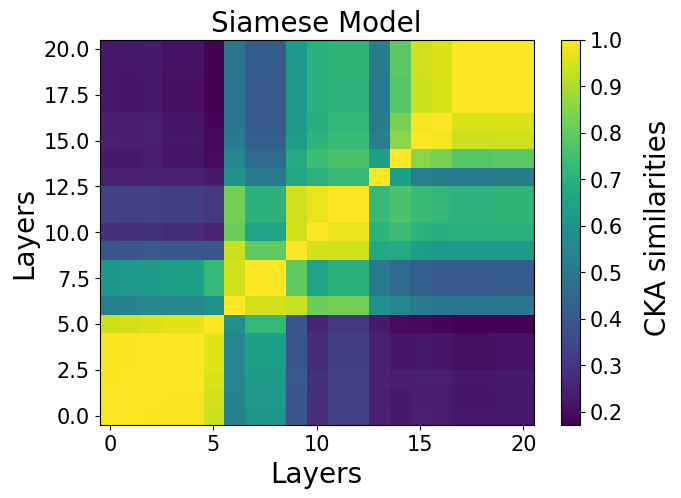}
    \caption{CKA similarities for the MLP (left),  CNN  (middle) and SNN  (right) models for the signal (BP with $m_A = 300$ GeV) using 1000 test events are shown. Layers  include the input, fully-connected, convolution, pooling  and dropout layers but exclude the output layer.}
\label{fig:1_11}
\end{figure}

 The Hilbert Schmidt Independent Criterion (HSIC) is defined as 
 \begin{equation}
     {\rm HSIC}(M,N) = \frac{1}{(d-1)^2}\ tr(MHNH)\,,
 \end{equation}
 with $H$ is a centering matrix and $M,N$ are defined above. It is worth mentioning that the HSIC is not invariant to random scaling of the input features, but it can be made invariant through a normalization as introduced in the CKA formula. The value of a CKA similarity ranges between $[0,1]$ and  indicates the similarity of the learned representations by each hidden layer. Layers with small CKA values do not share the same representations and they learn different information from the input data which improves the model classification performance. A larger CKA value indicates that  layers learn the same information about the input features resulting in no improvement of the classification accuracy of the model. In this case one can truncate those layers which share the same information to reduce the model complexity with no impact on the classification performance. For illustration of the relation between the CKA similarity of the hidden layers and the classification accuracy we point out to Fig. 3 in \cite{cka}. Here, Fig. \ref{fig:1_11} shows the CKA similarity for three DNN models: MLP (left plot), CNN (middle plot) and SNN (right plot). The models are trained on the signal point with $m_A = 300$ GeV and to compute the CKA we adopt $1000$ test samples for each model. The CKA value is then computed for all layers of each model, e.g., Conv2d, FC, dropout, pooling and input layers, but we do not include the final output layer (the two neurons layer with softmax activation).   
 The MLP model shows a uniform similarity distribution among all layers except the input layer. The uniform similarity in the MLP model indicates that the model is able to capture global information only. In fact, such a behavior  of the MLP layers is expected as the input features are high-level kinematic distributions which encode the global characteristic features of the signal and background events. Moreover, the uniform similarity among the MLP hidden layers indicates that one can reduce the number of the used layers with no significant reduction of the classification accuracy.

The CNN model shows a large CKA similarity among each convolution layer pair.  The first two convolution layers and the pooling layer have large CKA similarities. Also, the second and third pair of the convolution layers as well as the pooling layers have  large CKA similarities among themselves. The last layers, $14^\text{th} - 16^\text{th}$, are the fully-connected layers which have similar representation among themselves but are different from the convolution layers. In general, all convolution layers share a CKA similarity $\sim 0.8$ among each other which indicates that they all capture specific local information encoded into the jet images. 

As for the SNN model, when tested on the same CNN input, we notice that, while the first two convolution layers and the pooling layers have large CKA value as in the CNN model, the other convolution layers have small CKA similarity to the first convolution layers. The last layers, $15^\text{th} - 20^\text{th}$, are the fully-connected and dropout layers. The fact that the first couple of convolution layers do not share the same similarity to the later convolution layers indicates that convolution layers in the SNN capture different information from the input jet images. Indeed, the additional information captured by SNN  layers is the reason for the enhanced classification accuracy over the CNN one. Overall, the CKA similarity for SNN hidden layers assures that the convolution layers do not capture only local information (similar to the CNN layers) but also capture different types of information, similarity and dissimilarity of the input images. 

\begin{figure}[h!]
\centering
    \includegraphics[scale=0.45]{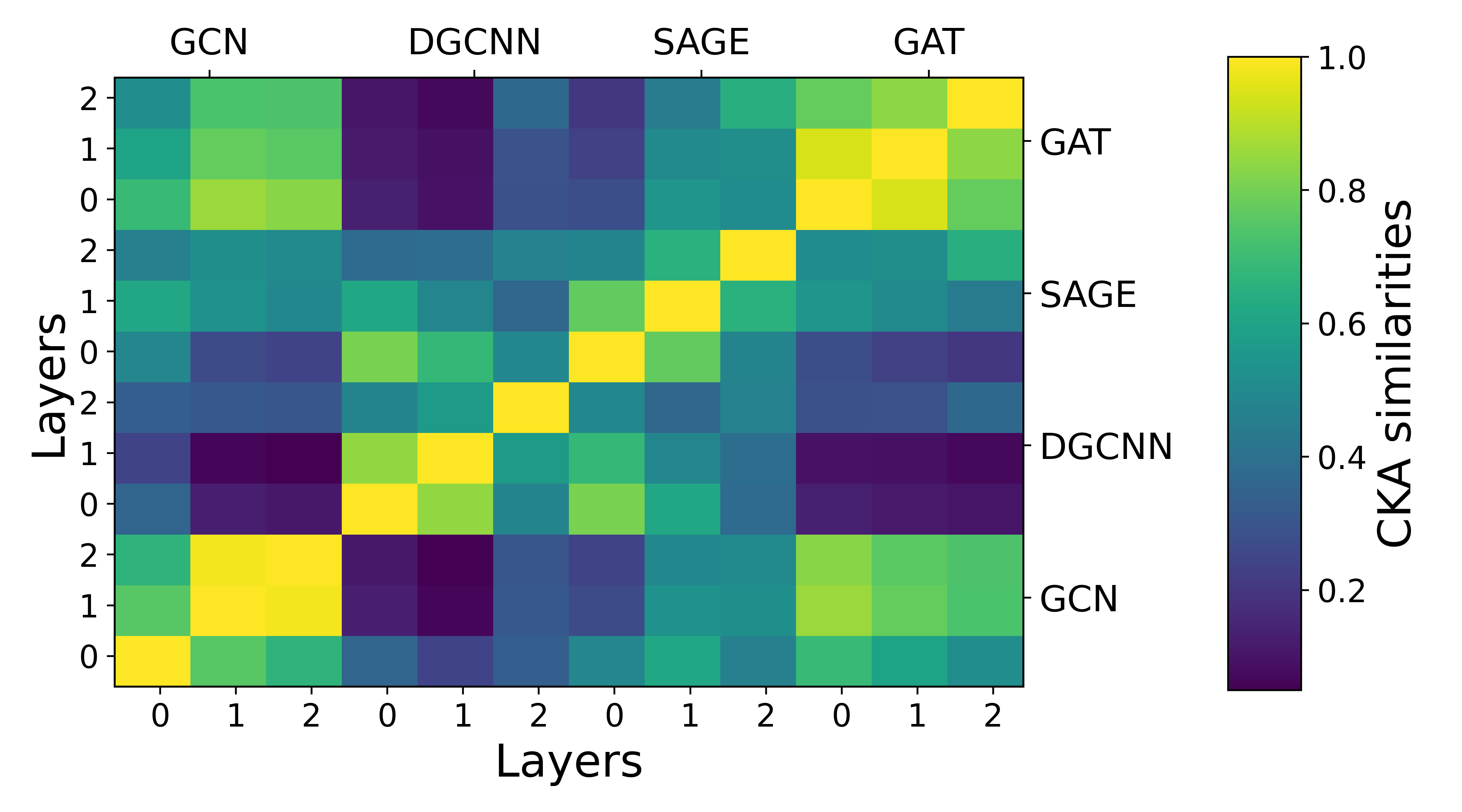}
    \caption{CKA Similarity for GNN models. The CKA values are computed for the layers of each model as well as layers from different models. }
\label{fig:1_12}
\end{figure}

The power of the CKA approach lies in its ability to compute the similarity between hidden layers from different models. As seen in Fig. \ref{fig:1_12}, we demonstrate the CKA similarity across all hidden layers of the GNN models. These similarities are drawn both within the hidden layers of each individual model and between the hidden layers from different GNN models.

An interesting observation is that the hidden layers from the DGCNN and GraphSAGE models demonstrate a relatively high similarity, approximately 0.7, but are distinctly different from the GCN and GAT models. This similarity reflects the classification accuracy values, with DGCNN and GraphSAGE displaying nearly identical classification accuracy across all mass points, as shown in Fig. \ref{fig:limits}.

\section{Results}
We now apply the different ML models to probe the $l^+l^-b\bar b$ ($l=e,\mu$) signature of the $pp\to A\to Z^{(*)}h$ process at Run 3 of the LHC (with an integrated luminosity of $300$ fb$^{-1}$) and  the HL-LHC (with an integrated luminosity of $3000$ fb$^{-1}$). The discrimination power of each of the networks measures how well the signal and background features are recognized,   which is quantified by the ROC curve. The better discrimination performance between signal and backgrounds, the higher  the true positive rate than the false positive rate in the ROC curve. Detailed information about the remaining number of signal and background events after optimizing the cuts on the DNN output for all the considered signal points can be found in Tab. \ref{tab:tablelabel}.  

\begin{figure}[h]
\centering
    \includegraphics[width=7.5cm,height=5cm]{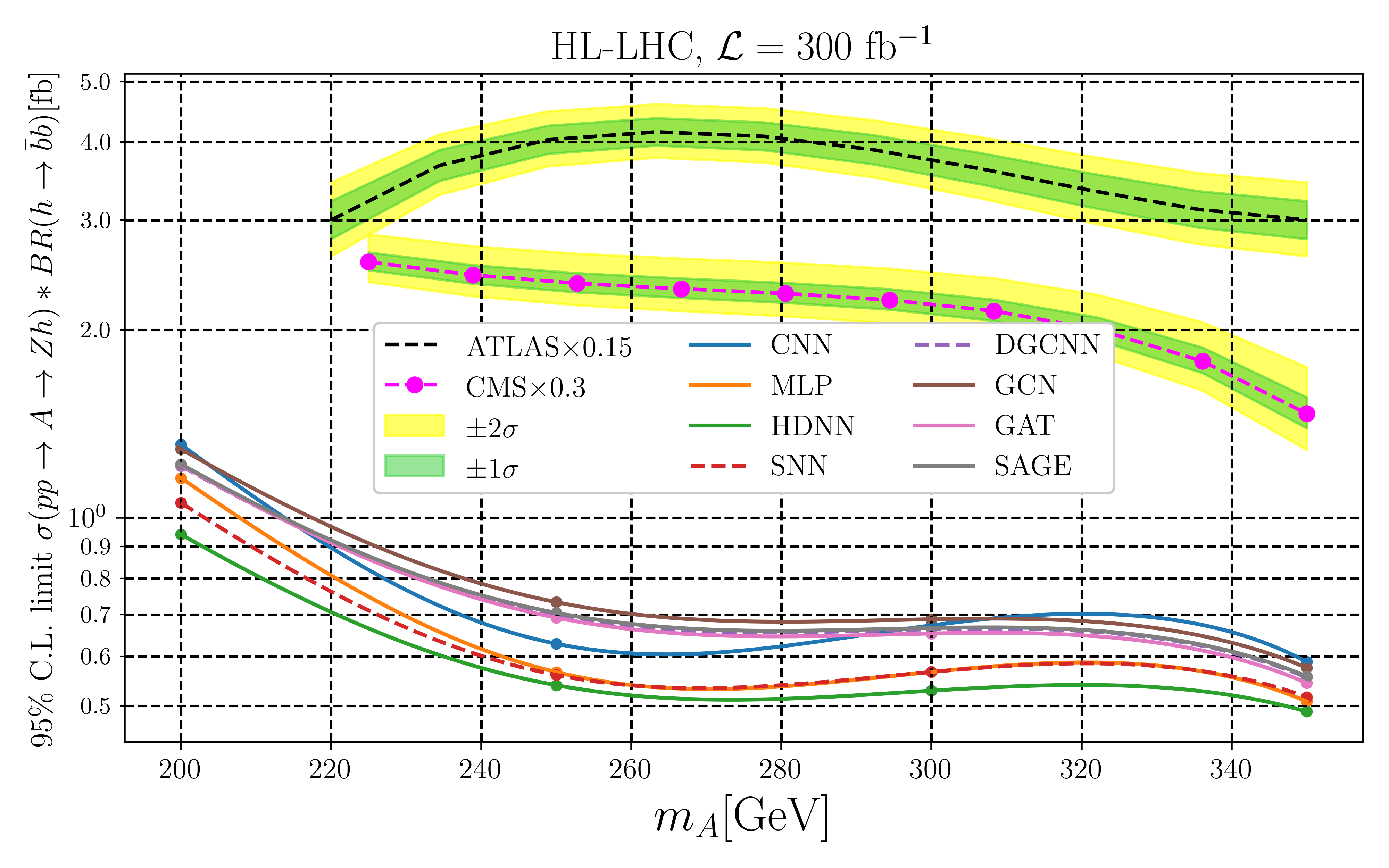}\includegraphics[width=7.5cm,height=5cm]{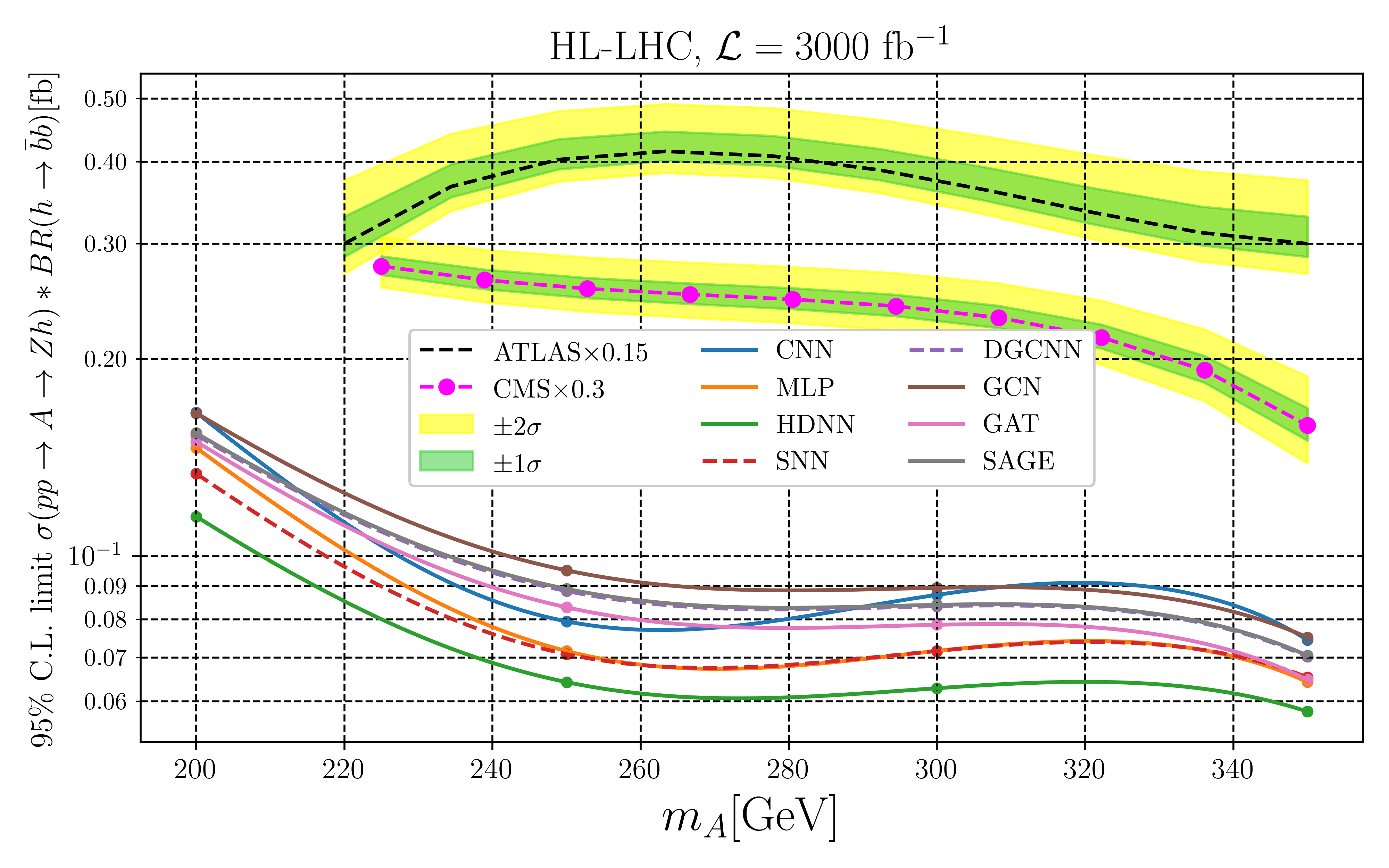}
    \caption{$95\%$ C.L. upper limit on the total cross section for the process $\sigma(pp\to A\to Zh)\times$ $Br(h\to\bar{b}b)$ at Run 3 of the LHC with $L_{\rm int}=300$ fb$^{-1}$ (left) and the HL-LHC with $L_{\rm int}=3000$ fb$^{-1}$ (right), both with $\sqrt{s}=14$ TeV and assuming a systematic uncertainty $\sigma_b=10\%$ on the background. The expected CMS and ATLAS limits extracted from \cite{CMS:2019qcx} (Fig. 5) and  \cite{TheATLAScollaboration:2016loc} (Fig. 11), respectively,  and linearly scaled to the two luminosities are also given. Bullet points on the ML curves indicate the four BPs considered. }
\label{fig:limits}
\end{figure}

The expected upper limit on the total cross section can be constructed by computing the probability of finding the expected data incompatible with the prediction of the various ML models. The expectation value of having a certain number of events in the $i^\text{th}$ $m_A$ bin in the DNN output score distribution is \cite{Cowan:2010js}
\begin{equation}
    E = \mu S_i +B_i\,,
\end{equation}
where $S_i$ and $B_i$ are the number of signal and background events, respectively, and $\mu$ is the signal strength parameter. The signal strength parameter defines the type of statistic measure,  so that $\mu=1$ is  rejecting a signal discovery hypothesis and defining the upper limit on the total cross section. Such a limit can be calculated from the optimization of the signal-to-background cut on the DNN output  and this has been done using the following significance formula \cite{LHCDarkMatterWorkingGroup:2018ufk,Antusch:2018bgr,Antusch:2020fyz}:
\begin{equation}
\sigma_{sys} = \left[ 2\left( (N_s+N_b)\ln\frac{(N_s+N_b)(N_b+\sigma^2_b)}{N_b^2+(N_s+N_b)\sigma^2_b}  -\frac{N^2_b}{\sigma^2_b}\ln(1+\frac{\sigma^2_b N_s}{N_b(N_b+\sigma^2_b)})         \right) \right]^{1/2}\,,
\end{equation}
with $N_s$ and $N_b$ being the number of signal and background events, respectively, and where $\sigma_b$ is the total uncertainty in the background events. For a $95\%$ Confidence Level (C.L.) upper limit on the total cross section for $\sigma(pp\to A\to Z^{(*)}h)\times$ $Br(h\to\bar{b}b)$, we require the signal significance to be $\sigma_{sys} \le 2$ \cite{LHCDarkMatterWorkingGroup:2018ufk}. The corresponding results for all considered ML models are shown in Fig. \ref{fig:limits}. The CMS and ATLAS bounds extracted from  \cite{CMS:2019qcx} (Fig. 5) and \cite{TheATLAScollaboration:2016loc} (Fig.  11),  respectively,  linearly scaled to the considered integrated luminosities, are also presented\footnote{It is worth mentioning here that the ATLAS analysis considers a combined limit between the channels of two isolated leptons plus no leptons. If, for consistency with our analysis,  one considered only the ATLAS results from the channel with two isolated leptons channel, the experimental  limit in Fig. \ref{fig:limits} will be relaxed.}. 


The overarching observation here is that the results obtained from all our ML approaches systematically outperform the experimental results for both LHC configurations. Amongst the former, as obviously seen, the HDNN with two stream inputs has the most  stringent limit  amongst all networks. Indeed, this is expected as merging both global and local information by concatenating an MLP and CNN enables the model to access all needed information about the events which in turns enhances the classification performance. Interestingly, the SNN has an equal, and even better, classification performance than the MLP network. Although the SNN  is trained on jet images, which encode only local information about the event, i.e., radiation patterns of the charged hadrons, it has more  stringent limits than the MLP network, which is trained on kinematic distributions encoding only global information. We interpret this as follows: that learning the similarity among events from the same class can secretly provide ``unseen'' global information to the SNN  model. This can be achieved by mapping events from different classes, signal and background, into different locations in the Euclidean latent space of the model during the first training stage. The newly learned ``unseen'' global information can be clearly interpreted from the CKA of SNN where the early convolution layers capture information different to the ones captured by the late convolution ones. 
{Moreover, GNNs (DGCNN, GCN, GAT and GraphSAGE) produce a  slightly worse performance than CNNs. This is because they analyze low level data, i.e., (reconstructed) four-momenta of the final state particles. To increase the classification performance of GNN models, we would need to add more inductive biases to be tuned. In fact, the size of the constructed graphs is small, which is due to the small number of (reconstructed) final state particles in each event. Accordingly, if we increased the number of the inductive biases, e.g., by increasing the number of GNN layers, the GNN model would overfit}.  

\begin{table}[h!]
\caption{Analysis summary for all DNN models considered (column 1) for all signal BPs from Tab.~\ref{tab:tab_1}, identified by the $A$ mass (column 2). Column 3 shows the area under the ROC curve in each case. Columns 5 and 6 show the number of remaining signal and background events, respectively, after maximizing the cut on the DNN output. Column 6 shows the final significance. For illustrative purposes, event rates and significances are here computed at the luminosity mid point of 1000 fb$^{-1}$.}\vspace*{0.25cm}\hspace*{0.65cm}
\label{tab:tablelabel}
   \begin{tabular}{|c|c|c|c|c|c|} 
\cline{2-6}
    \multicolumn{1}{c|}{}&BPs& AUC & Signal ($S$) & Background ($B$) & $\sigma_{}$\\
    \hline
    \multirow{4}{*}{MLP} & $m_A=200$ GeV & 0.90 & 6538 & 486 & 156 \\ \cdashline{2-6}[.4pt/1pt]
                         & $m_A=250$ GeV& 0.92 & 46894 & 1422 & 496 \\ \cdashline{2-6}[.4pt/1pt]
                         & $m_A=300$ GeV& $0.95$ & 63060 & 1287 &614 \\ \cdashline{2-6}[.4pt/1pt]
                         & $m_A=350$ GeV& 0.96 & 69496 & 1004 & 678 \\ \cdashline{2-6}[.4pt/1pt]
    
    \hline
    \multirow{4}{*}{CNN}& $m_A=200$ GeV& 0.87 & 9253 &2007 & 142 \\ \cdashline{2-6}[.4pt/1pt]
                        & $m_A=250$ GeV& 0.89 & 47122 &2675 & 443 \\ \cdashline{2-6}[.4pt/1pt]
                        & $m_A=300$ GeV& $0.86$ & 60114 &4417 & 485 \\ \cdashline{2-6}[.4pt/1pt]
                        & $m_A=350$ GeV& 0.90 & 66320 &2364 & 574 \\ \cdashline{2-6}[.4pt/1pt]
    
    \hline
    \multirow{4}{*}{SNN}& $m_A=200$ GeV& 0.92 & 7267 &441 & 171 \\ \cdashline{2-6}[.4pt/1pt]
                        & $m_A=250$ GeV& 0.94 & 45822 &1135 & 507 \\ \cdashline{2-6}[.4pt/1pt]
                        & $m_A=300$ GeV& $0.95$ & 61940 &1210 & 612 \\ \cdashline{2-6}[.4pt/1pt]
                        & $m_A=350$ GeV& 0.96 & 69894 &2843& 672 \\ \cdashline{2-6}[.4pt/1pt]
    
    \hline
    \multirow{4}{*}{HDNN}& $m_A=200$ GeV& 0.93 & 7810 &328 & 191 \\ \cdashline{2-6}[.4pt/1pt]
                        & $m_A=250$ GeV& 0.95 & 44033 &767 & 525 \\ \cdashline{2-6}[.4pt/1pt]
                        & $m_A=300$ GeV& $0.98$ & 63243 &784 & 661 \\ \cdashline{2-6}[.4pt/1pt]
                        & $m_A=350$ GeV& 0.97 & 69798 &956 & 685 \\ \cdashline{2-6}[.4pt/1pt]
    
    \hline
    \multirow{4}{*}{DGCNN}& $m_A=200$ GeV& 0.86 &6973 & 722 & 150 \\ \cdashline{2-6}[.4pt/1pt]
                        & $m_A=250$ GeV&  0.83&39126&1958 & 414 \\ \cdashline{2-6}[.4pt/1pt]
                        & $m_A=300$ GeV& $0.87$ &51436 &1362 &532 \\ \cdashline{2-6}[.4pt/1pt]
                        & $m_A=350$ GeV& 0.91 & 63463 &1131 & 608 \\ \cdashline{2-6}[.4pt/1pt]
    
    \hline
    \multirow{4}{*}{GCN}& $m_A=200$ GeV& 0.84 & 7256 &992 & 143\\ \cdashline{2-6}[.4pt/1pt]
                        & $m_A=250$ GeV& 0.78 &38242 &2083 & 403 \\ \cdashline{2-6}[.4pt/1pt]
                        & $m_A=300$ GeV& $0.84$ & 50154 &1416 & 520 \\ \cdashline{2-6}[.4pt/1pt]
                        & $m_A=350$ GeV& 0.89 & 61510 &1051& 620 \\ \cdashline{2-6}[.4pt/1pt]
    
    \hline
    \multirow{4}{*}{GAT}& $m_A=200$ GeV& 0.85 &6904 &745 &147 \\ \cdashline{2-6}[.4pt/1pt]
                        & $m_A=250$ GeV& 0.82& 38459 &1894 & 412 \\ \cdashline{2-6}[.4pt/1pt]
                        & $m_A=300$ GeV& $0.85$ & 51237 &1407 & 528 \\ \cdashline{2-6}[.4pt/1pt]
                        & $m_A=350$ GeV& 0.9 & 63607 &1223 & 622\\ \cdashline{2-6}[.4pt/1pt]
    
    \hline
    \multirow{4}{*}{GraphSAGE}& $m_A=200$ GeV& 0.85 & 6915 &723 & 148 \\ \cdashline{2-6}[.4pt/1pt]
                        & $m_A=250$ GeV& 0.82 & 38993 &1994& 412 \\ \cdashline{2-6}[.4pt/1pt]
                        & $m_A=300$ GeV& $0.86$ &51284 &1392 & 529 \\ \cdashline{2-6}[.4pt/1pt]
                        & $m_A=350$ GeV& 0.9 & 63655 &1207 & 624 \\ \cdashline{2-6}[.4pt/1pt]
    
    \hline
\end{tabular}
\end{table}

\section{Conclusions}
\label{sec:intro}
In summary, in this paper, we have used a variety of advanced ML techniques, most of which had never been applied previously to BSM Higgs boson searches (although various graph and transformer networks have been exploited for QCD studies related to jet dynamics), to prove that they can offer a significant improvement with respect to traditional LHC analyses, using either a cut-and-count approach or else based on traditional ML approaches, such as (shallow) NNs. Our methodology, in fact, exploits instead DNNs, the latter covering MLP, CNN, SNN and HDNN algorithms and a variety of GNNs (DGCNNs, GCNs, GAT and GraphSAGE networks), including an interpretable ML element that gives us confidence in the accuracy of our predictions. 

In order to exemplify the scope of this multi-prong ML approach, we have targeted a BSM process to which current experimental analyses by ATLAS and CMS have moderate sensitivity, i.e., $b\bar b,gg\to A\to Z^{(*)}h\to l^+l^-b\bar b$ ($l=e,\mu$), involving the production and decay of a CP-odd Higgs state of a 2HDM Type-II ($A$) and the SM-like Higgs discovered in 2012 ($h$). The CERN machine configurations adopted included Run 3 of the LHC as well as the HL-LHC, wherein $\sqrt{s}=14$ TeV for both and $L_{\rm int}=$ 300 and 3000 fb$^{-1}$, respectively.  This process is rather intriguing, as it is a potential BSM signal that would prove simultaneously the presence of an extended Higgs sector (with a different CP nature) with respect to the SM one and its underlying gauge structure. However, such a BSM signal suffers from overwhelming backgrounds, so that it is a significant challenge to extract it. Furthermore, depending on the $A$ mass, whether it is such that 
$m_A<m_Z+m_h$, $m_Z+m_h<m_A<2m_t$ or $m_A\approx2m_t$, both the size of the signal (via the  $Br(A\to Z^{(*)})$, with the gauge boson being either off- or on-shell as $m_A$ increases) and the composition of the background samples vary significantly, so that different kinematical selections are generally required to optimise the sensitivity of the various searches therein. 

By adopting standard acceptance cuts on the final state objects (leptons and hadrons) and a rather bland selection around the $Z^{(*)}$ and $h$ masses, so long that these are supplemented by a  combination of the aforementioned ML tools, we were able to improve, in comparison to the very latest ATLAS and CMS results,  the sensitivity to the cross section of the signal process by at least a factor of 4(2) over the $m_Z+m_h<m_A<2m_t$($m_A\approx 2m_t$)  region while at the same time proving that there can be sensitivity also over the so-far unexplored $m_A<m_Z+m_h$ interval. 

These improved results, obtained generally from various advanced ML approaches, are most noticeable when concatenating a MLP and CNN into a two stream HDNN model, which is then the approach we would recommend to establish the aforementioned BSM Higgs signal.

\section*{Acknowledgments}
AH thanks Mihoko Nojiri for the fruitful discussion about the SSN.
AH is funded by the grant number 22H05113,"Foundation of "Machine Learning Physics", Grant in Aid for Transformative Research Areas and 22K03626, Grant-in-Aid  for Scientific Reseach (C). SM is supported in part through the NExT Institute and the STFC Consolidated Grant No. ST/L000296/1.
\appendix

\section{Evaluation metrics for a BP }

\begin{figure}[th!]
   \centering
    \includegraphics[scale=0.2]{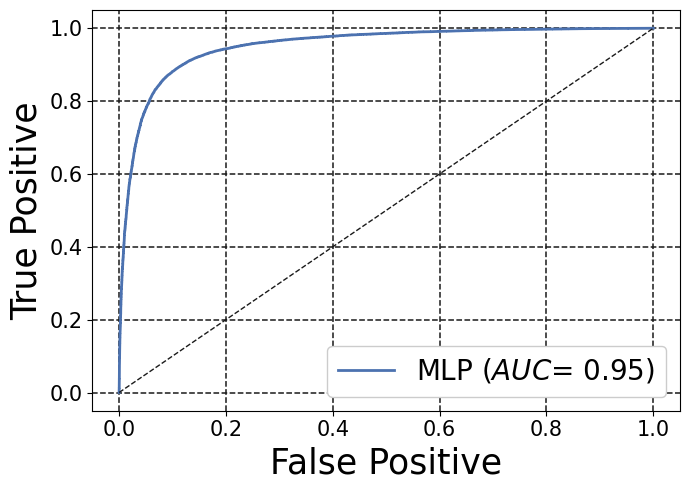}~~
    \includegraphics[scale=0.2]{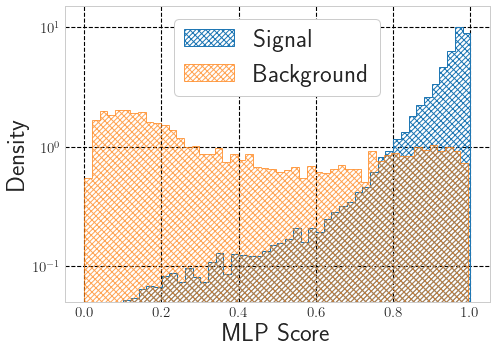}~~
    \includegraphics[scale=0.233]{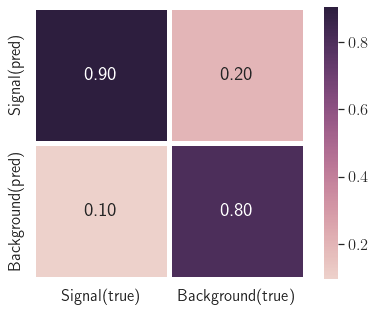} 
    \\
    \includegraphics[scale=0.2]{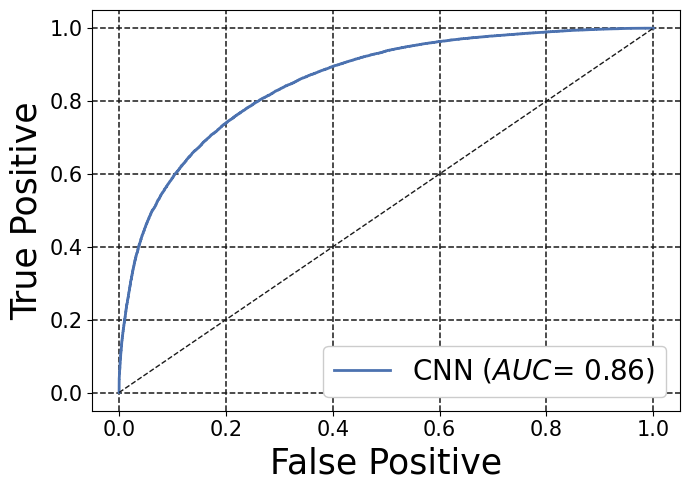}~~
    \includegraphics[scale=0.2]{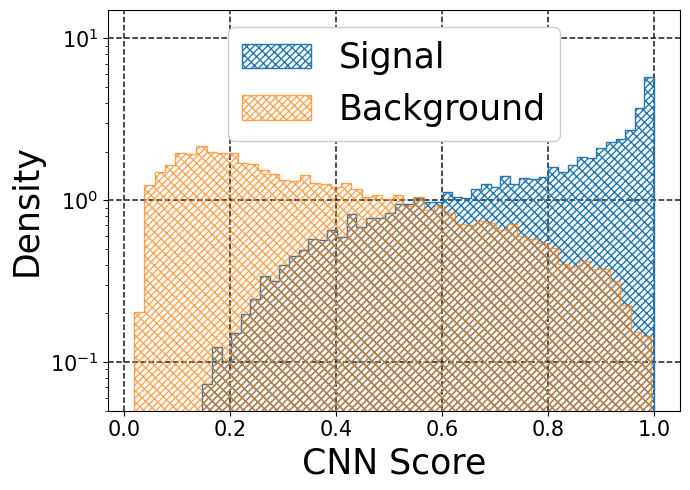}~~
    \includegraphics[scale=0.233]{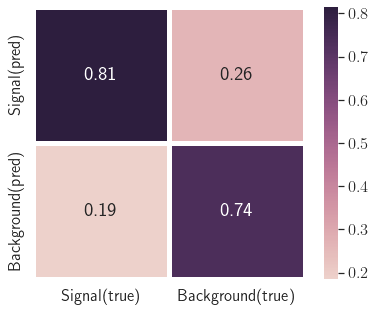}
    \\
    \includegraphics[scale=0.2]{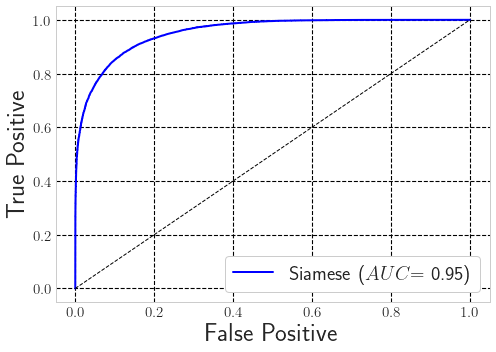}~~
    \includegraphics[scale=0.2]{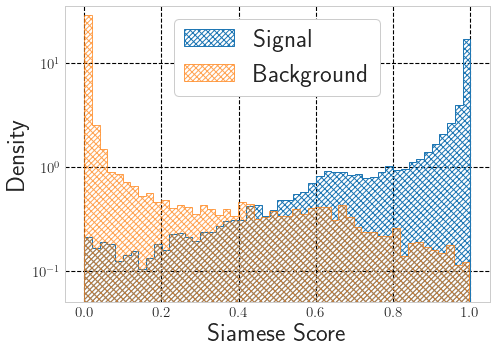}~~
    \includegraphics[scale=0.233]{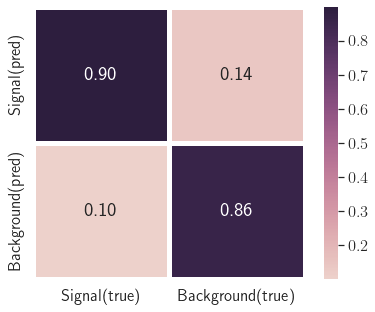}
    \\
    \includegraphics[scale=0.2]{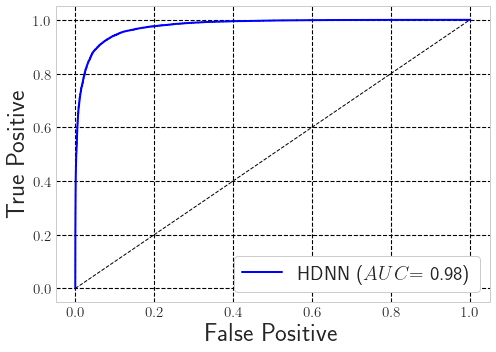}~~
    \includegraphics[scale=0.2]{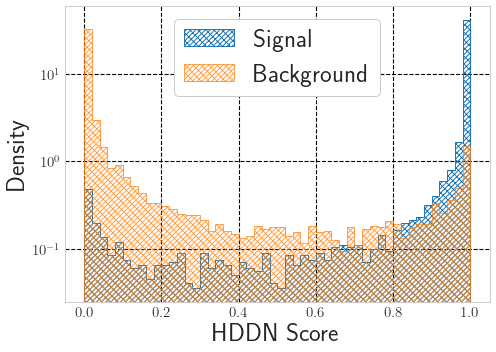}~~
    \includegraphics[scale=0.233]{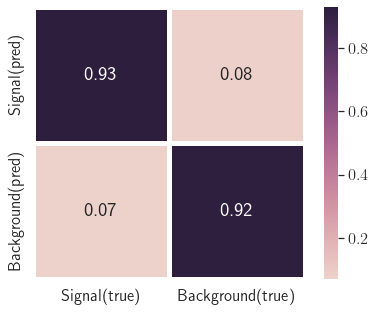}
    \\
    \includegraphics[scale=0.2]{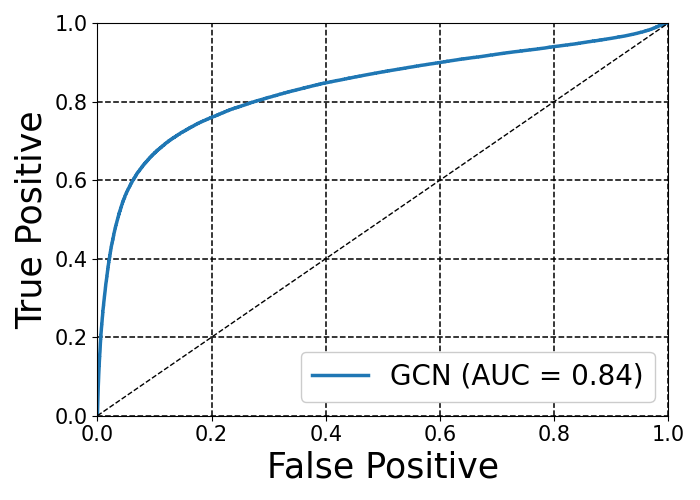}~~
    \includegraphics[scale=0.2]{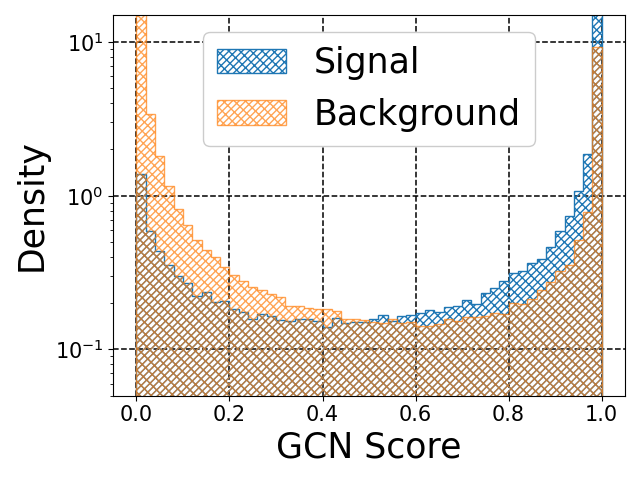}~~
    \includegraphics[scale=0.2]{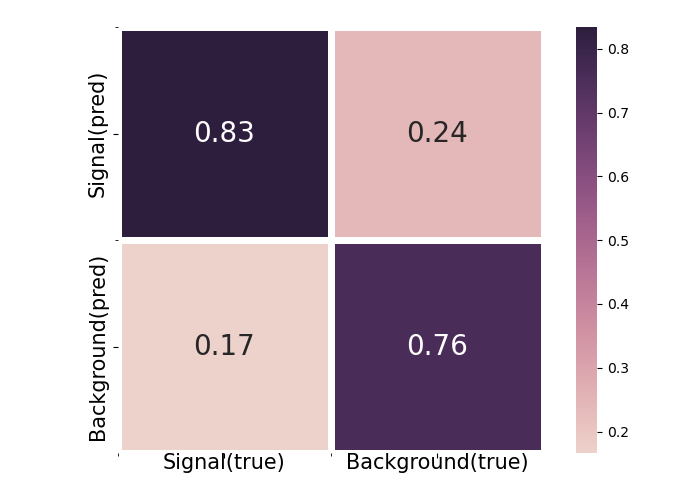}
    \\
    \includegraphics[scale=0.2]{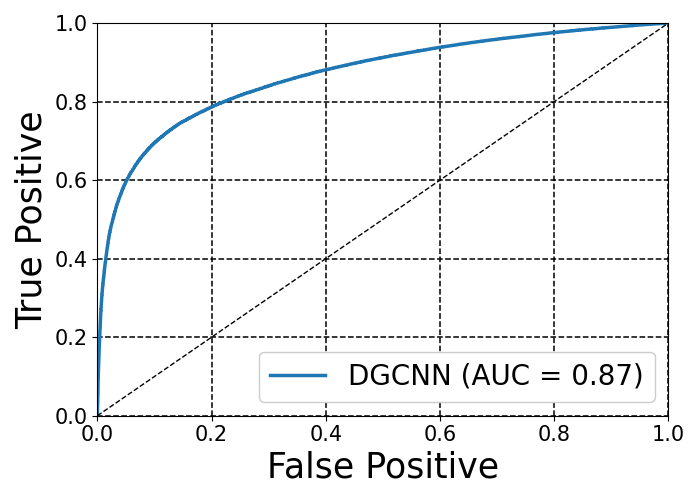}~~
    \includegraphics[scale=0.2]{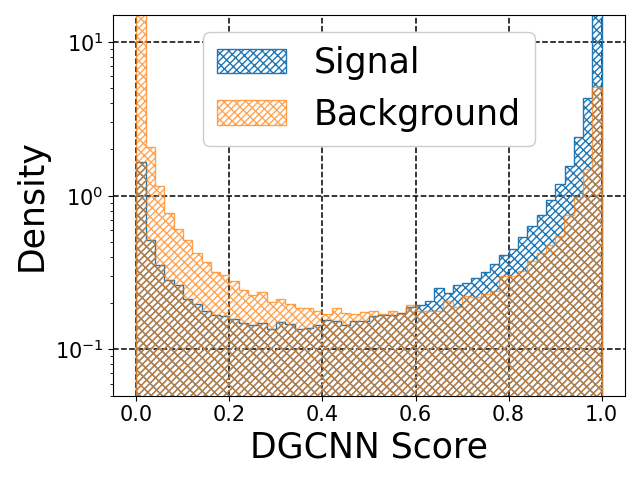}~~
    \includegraphics[scale=0.2]{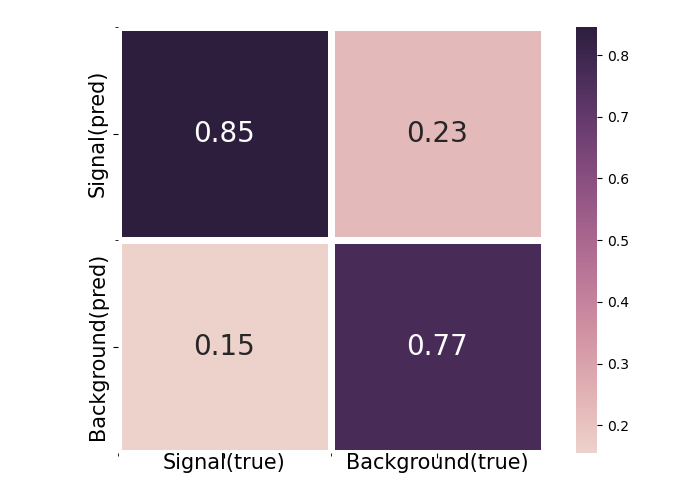}
    \\
    \includegraphics[scale=0.2]{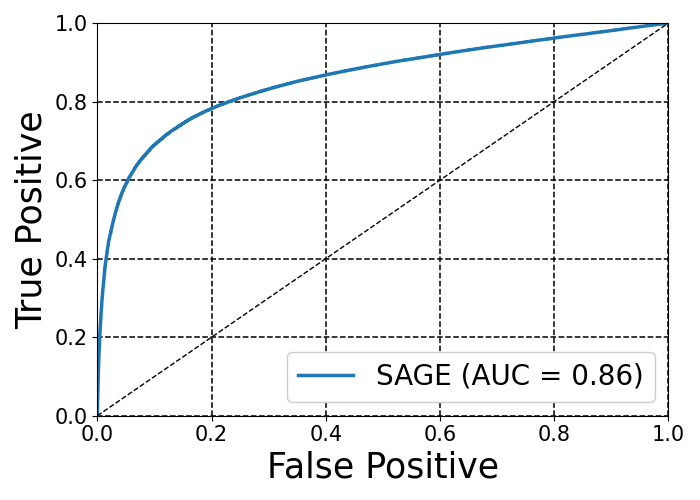}~~
    \includegraphics[scale=0.2]{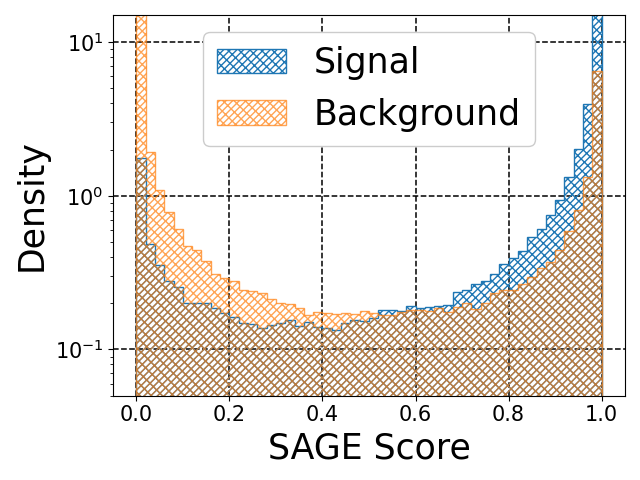}~~
    \includegraphics[scale=0.2]{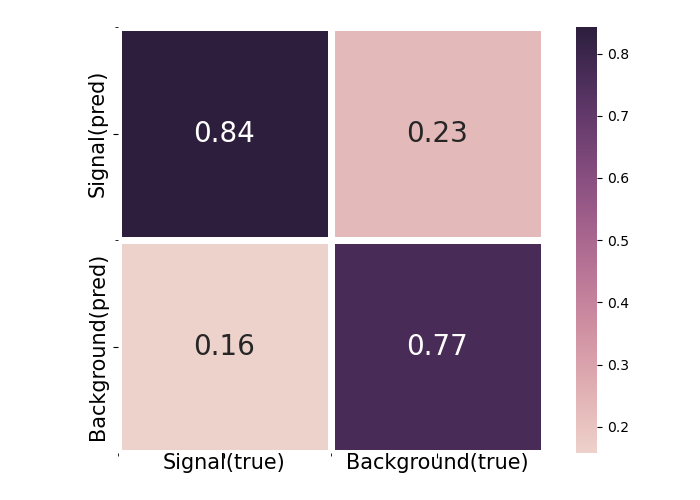}
    \\
    \includegraphics[scale=0.2]{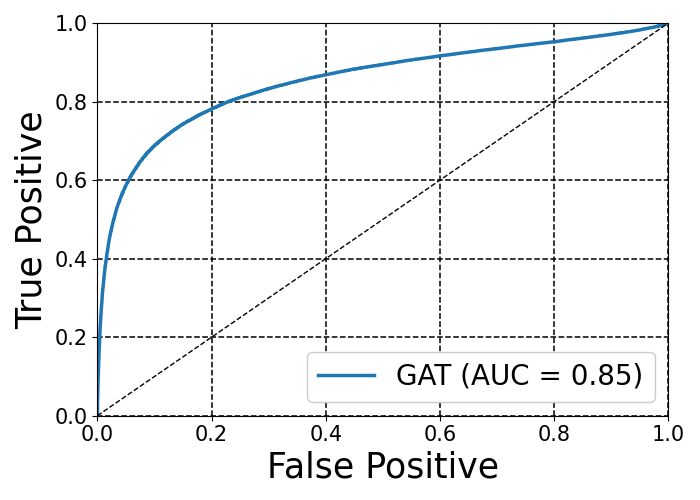}~~
    \includegraphics[scale=0.2]{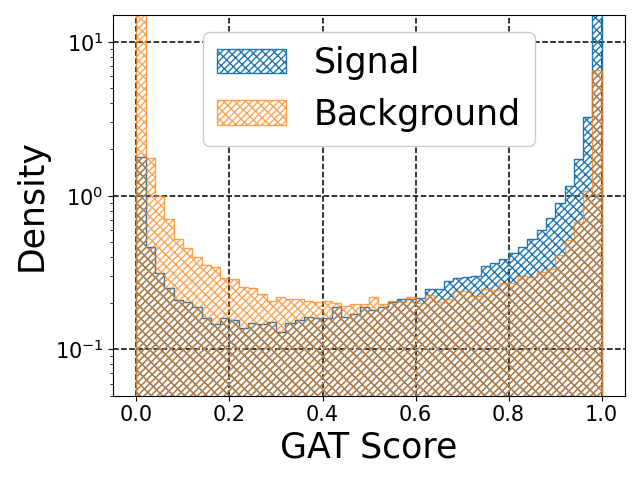}~~
    \includegraphics[scale=0.2]{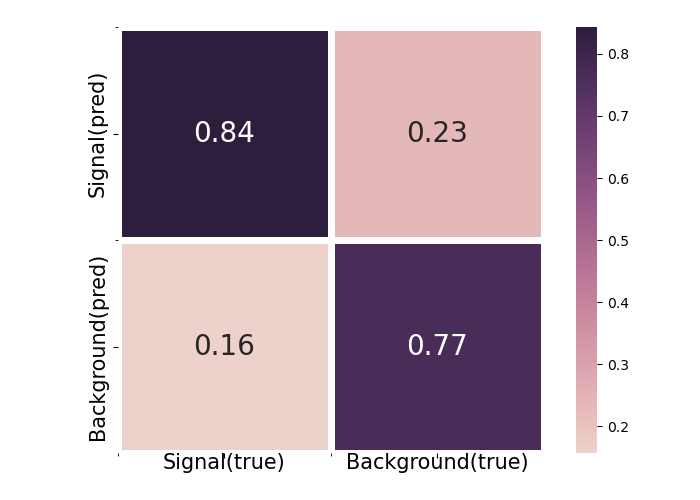}

    \caption{Test results when trained on kinematic distributions for signal events (BP with $m_A=300$ GeV). 
    The ROC curve (left),  output score (middle) and confusion matrix (right) for all the considered DL models.}
\label{fig:1_3}
\end{figure}

In this appendix we show the results for the used benchmark point with $m_A= 300$ GeV. 
The ROC curve is an evaluation metric for binary classification problems: it is a probability curve that plots the true positive rate  against the false positive rate at various threshold values and essentially separates the `signal' from the `noise' caused  by misclassifying the background. In other words, it shows the performance of the model to identify the signal events at all classification thresholds. The Area Under the Curve (AUC) is the measure of the ability of a binary classifier to distinguish between classes and is used as a summary of the ROC curve. As the ROC curve quantifies the relation between true and false positive rates, which indicates the model ability to identify the signal events, the Confusion Matrix (CM) reports the values for both positive and negative hypotheses. Accordingly, one can clearly estimate the model response to identify the signal and background events. 
The output score of  the considered DLs has  the dimension of  $1\times 2$, ($\mathcal{P}_\text{sig},\mathcal{P}_\text{bkg}$), with $\mathcal{P}$ ranges between $[0,1]$. If $\mathcal{P}_\text{sig} > 0.5$ the corresponding event is classified as most likely being a signal event and if $\mathcal{P}_\text{bkg} > 0.5$ the corresponding event is classified as most likely a background event. Output score plots represent $(\mathcal{P}_\text{sig},1-\mathcal{P}_\text{bkg})$ with events near 1 are classified as most likely signal events and events near 0 are classified as most likely being background events.


\end{document}